\documentclass[12pt,nofootinbib, floatfix, amsmath, amssymb,aps,prd,preprint]{revtex4-1}
\usepackage[T1]{fontenc}
\usepackage{amsmath, amssymb, tabularx}
\usepackage{lipsum}
\newcommand{\tagarray}{%
\mbox{}\refstepcounter{equation}%
$(\theequation)$%
}
\usepackage{color}
\usepackage{colordvi}
\usepackage{amsmath}
\usepackage{amssymb}
\usepackage{amsfonts}
\usepackage{graphicx}
\usepackage{multirow}
\usepackage[dvipsnames]{xcolor}
\usepackage[colorlinks=true,urlcolor=blue,linkcolor=blue,citecolor=blue]{hyperref}
\newcommand\pdfmath[1]{\texorpdfstring{$#1$}{#1}}
\usepackage{cancel}
\usepackage{caption}
\usepackage{subcaption}
\usepackage{float}
\usepackage{gensymb}
\usepackage{soul}
\usepackage[top=1.135in,bottom=1.17in,left=1.16in,right=1.16in]{geometry}

\usepackage{silence}
\allowdisplaybreaks

\newcommand{\eq}[1]{\begin{align}#1\end{align}}

\newcommand{\hc}{{\rm h.c.}}

\usepackage[dvipsnames]{xcolor}
\usepackage{ulem}

\begin{document}
\thispagestyle{empty}

\title{Non-unitary limits on different textures for low-scale seesaw models. }

\vspace{50pt}
\author{Jes\'us Miguel Celestino-Ram\'irez$^{1}$\footnote{jesus.celestino@cinvestav.mx}, G. Hern\'andez-Tom\'e$^{2}$\footnote{gerardo\_hernandez@uaeh.edu.mx}, O. G. Miranda$^{1}$\footnote{omar.miranda@cinvestav.mx}, Eduardo Peinado$^{1,3}$\footnote{epeinado@fisica.unam.mx}}\affiliation{$^1$Departamento de F\'isica, Centro de Investigaci\'on y de Estudios Avanzados del Instituto Polit\'ecnico Nacional\\
Apartado Postal 14-740, 07000 Ciudad de M\'exico, M\'exico\\
$^2$Área Académica de Matemáticas y Física, Universidad Autónoma del Estado de Hidalgo, Carretera Pachuca-Tulancingo Km. 4.5, C.P. 42184, Pachuca, Hgo.\\
$^3$Instituto de F\'isica, Universidad Nacional Aut\'onoma de M\'exico, AP 20-364, Ciudad de M\'exico 01000, M\'exico
}

\begin{abstract}

New heavy neutral leptons lead to non-unitary effects in models for neutrino masses. Such effects could represent a sign of new physics beyond the Standard Model, leading to observable deviations in neutrino oscillation experiments, lepton flavor violation, and other precision measurements. This work explores the parameter space of the linear and inverse low-scale seesaw models based on flavor symmetries consistent with neutrino oscillation experiments. In particular, we investigated the violation of unitarity when the lepton flavor violation is absent and when only one lepton flavor-violating channel is present.

\end{abstract}
\maketitle
\section{Introduction}

Whether neutrinos are Dirac or Majorana particles and what mechanism suppresses their masses are two open questions that are more urgent with the discovery of neutrino oscillation. One potential explanation for the latter is the seesaw mechanism \cite{Schechter:1981cv,Mohapatra:1986bd, Minkowski:1977sc,Yanagida:1979as,Foot:1988aq,Cheng:1980qt}. This work will focus on the low-scale type-I seesaw mechanism, specifically the linear ~\cite{Malinsky:2005bi} and inverse models ~\cite{Gonzalez-Garcia:1988okv}  \footnote{For readers interested in a thorough analysis of the Type-I Seesaw family of neutrino models, see reference \cite{CentellesChulia:2024uzv}.}. Contrary to the canonical high-scale Type-I seesaw, these scenarios offer an attractive alternative capable of explaining the tiny mass of neutrinos without requiring extremely high-energy scales, providing a rich phenomenological landscape testable with current and future experiments. For example, through sizable charged Lepton Flavor violation (cLFV) processes due to the new Neutral Heavy leptons (NHL) with masses at the TeV scale.

Nevertheless, as pointed out in reference \cite{Garnica:2023ccx}, it is worth noting that cLFV can also be suppressed in low-scale seesaw scenarios if specific structures in the Yukawa couplings are considered. Therefore, in this work, we study the phenomenology and explore the allowed parameter space regions of some possible low-scale seesaw realizations. Firstly, continuing with the study from reference \cite{Garnica:2023ccx}, we consider the case where any of the three $\mu-e$, $\tau-e$ or $\tau-\mu$ possible cLFV transitions are strongly suppressed, but accounting for the non-unitary effects due to the presence of the NHL. Moreover, we also discuss low-scale scenarios that allowed just one of the three possible cLFV transitions. This approach is similar to the one presented in reference \cite{Arganda:2014dta}, looking for low-scale realizations that could maximize cLFV processes compatible with neutrino oscillation data and the current constraints from the search of cLFV.

We provide models based on symmetries justifying the specific textures for the neutrino mass matrix, constraining their allowed parameter space regions by considering the non-unitarity limits coming from neutrino oscillation experiments \cite{deSalas:2020pgw}. These constraints are obtained using the symmetric parametrization \cite{Escrihuela:2015wra, Celestino-Ramirez:2023zox, Chatterjee:2021xyu, CentellesChulia:2024sff}, therefore, we will establish a correspondence between the symmetric and the $\eta$ parametrization that came from the see-saw mechanism. In this regard, several works have studied the phenomenology of non-unitarity in the context of all cLFV processes \cite{Garnica:2023ccx, Celestino-Ramirez:2023zox, CentellesChulia:2024uzv, Forero:2011pc, Batra:2023mds}.

The specific models we provide give rise to light neutrino mass matrices with two-zero textures ~\cite{Frampton:2002yf} that are not only compatible with the current neutrino oscillation experiments but are also predictive~\cite{Ludl:2011vv,Meloni:2014yea,DeLaVega:2018bkp,Alcaide:2018vni}.

The structure of the manuscript is as follows: In Section \ref{TI}, the conventional type-I seesaw mechanism and its low-scale variants are introduced. Section \ref{MODELS} introduces specific low-scale seesaw realizations based on flavor symmetries. Then, section  \ref{NA} discusses the phenomenology and the numerical analysis of these scenarios. Finally, in section 
 \ref{Conclusions}. we present a summary and the conclusions. 
 
\section{Low scale type-I seesaw models}\label{TI}

If neutrinos are Majorana particles, the dimension-5 Weinberg operator \cite{Weinberg:1979sa} $\mathcal{O}_W=(L\cdot H) (L\cdot H)/\Lambda$  provides a natural mechanism for generating small neutrino masses whose most popular UV completion is the the seesaw mechanism. After the Electroweak Symmetry Breaking (EWSB), the dimension-5 Weinberg operator gives rise to a Majorana mass term for neutrinos given by $m_\nu=v^2/\Lambda$, where  $v=246$ GeV is the SM Higgs {\it vev} and $\Lambda$ is a high-energy scale at which new physics becomes relevant.

The type-I seesaw consists of adding right-handed (RH) sterile neutrinos $N_i$. Although elegant and appealing from a theoretical perspective, the type-I seesaw model inherently leads to highly suppressed phenomenological effects owing to the tiny seesaw expansion parameter $\epsilon^2=\mathcal{O}(m_\nu/M)$.

In contrast, low-scale variants of the model, such as the inverse and linear models, present a phenomenology that is potentially accessible at current or near-future collider experiments. These scenarios require incorporating new right-hand (RH) and left-handed (LH) sterile neutrinos $S_i$ typically around the TeV scale or lower, making their phenomenology accessible at current or near-future collider experiments. In this work, we consider the following Lagrangian with three pairs of N and S states
\eq{
{\cal L}_{Yuk}
=
y^{(\ell)}\, \overline{\hat{L}}\, H\, {\hat{\ell}}_{R} + 
y^{(\nu)}\, \overline{\hat{L}}\, \tilde{H}\, \hat{N}_{R} + 
\tilde{y}^{(\nu)}\, \overline{\hat{L}}\, \tilde{H^{\prime}}\, \hat{S}^c +
M^{(N)}\, \overline{\hat{S}}\, \hat{N}_{R}\,+ \mu \overline{\hat{S}}\hat{S}^c
+\hc
\label{seesawgen}
}
where 
\eq{
H=\begin{pmatrix}H^+\\ H^0 \end{pmatrix},\quad H^{\prime}=\begin{pmatrix}H^{\prime+}\\ H^{\prime 0} \end{pmatrix},\quad L_{i}=\begin{pmatrix}\nu_{Li}\\ \ell_i \end{pmatrix},}
 $\tilde{H}=i\sigma_2 H^*$ and $\tilde{H^{\prime}}=i\sigma_2 H^{\prime*}$. In such a way that when the matrix $\tilde{y}^{(\nu)}$ is absent, we refer to the model as the inverse seesaw, whereas when the matrix $\mu$ vanishes, we call it the linear seesaw. 

The global symmetry $U(1)_{B-L}$ with three RH neutrinos is anomaly-free,  meaning it can be gauged.  The linear see-saw (it also applies to the inverse see-saw) includes extra left-handed sterile states $S_i$,  remaining anomaly-free if the $S_i$ fields are $B-L$ singlets~\cite{Garnica:2023ccx}.

\section{Model for linear seesaw without cLFV}\label{MODELS} 

Consider the linear seesaw with three RH neutrinos whose $B-L$ charge is $-1$ and three LH sterile neutrinos with no $B-L$ charge. In this case, the theory is anomaly-free, the $B-L$ symmetry forbids the Majorana mass term for the RH neutrinos. To forbid the Majorana mass term for the LH sterile neutrinos and generate the diagonal structures for $M_L$, $M_D$, and $M_R$, we include a discrete flavor symmetry $Z_N$. All the fermions must transform non-trivially under $Z_N$. Assigning $f_1\sim \omega^1$, $f_2\sim \omega^2$, $f_3\sim \omega^3$ with $\omega^N=1$ where $f_i=L_i,N_i,S_i$. A condition to forbid any Majorana term is $N>6$. Consider the model in Table~\ref{tab:charges}.

\begin{table}[t]
    \centering
    \begin{tabular}{|c|c|c|c|c|c|c|c|c|c|c|c|c|c|c|c|}\hline
              &$L_e$ &$L_\mu$ &$L_\tau$ & $l_e$ & $l_\mu$   &$l_\tau$  & $N_1$ & $N_2$ & $N_3$ &$S_1$ & $S_2$ & $S_3$ & $H$&$H_i^\prime$ & $\phi$ \\ \hline
     $SU(2)_L$& 2 & 2 & 2 & 1 & 1 & 1 & 1 & 1 & 1 & 1 & 1 & 1 & 2 & 2 & 0\\ \hline
      $U(1)_{B-L}$ & $-1$ & $-1$ & $-1$ & $-1$ & $-1$ & $-1$ & $-1$ & $-1$ & $-1$ &$0$ & $0$ & $0$ & 0 &1&1 \\ \hline 
      $Z_7$ &  $\omega$ & $\omega^2$ & $\omega^3$ & $\omega$ & $\omega^2$ & $\omega^3$ & $\omega$ & $\omega^2$ & $\omega^3$ &$\omega$ & $\omega^2$ & $\omega^3$ & 1 &$\omega,~\omega^2,~\omega^3$ &1\\ \hline
    \end{tabular}
    \caption{\footnotesize $U(1)^{\prime}$ charges of the linear seesaw model. The charges $x_\alpha$, with $\alpha=e,\mu,\tau$, can take the values $x_\alpha = 0,-1,-2$, while the charges of the quarks are $1/3$.}
    \label{tab:charges}
\end{table}
 The Lagrangian in Eq. (\ref{seesawgen}) takes the form
\begin{gather*}
{\cal L}_{Yuk}\textit
=
y^{(\ell)}_i\, \overline{\hat{L}_{i}}\, H\, {\hat{\ell}}_{Ri} + 
y^{(\nu)}_i\, \overline{\hat{L}_{i}}\, \tilde{H}\, \hat{N}_{Ri} + 
\tilde{y}^{(\nu)}_i\, \overline{\hat{L}_{i}}\, \tilde{H}\, \hat{S}^c_{i} +
M^{(N)}_i\, \overline{\hat{S}_{i}}\, \hat{N}_{Ri}\,+
\hc
\label{lagrangian}
\end{gather*}
The structure of the $M_L$ matrix is 
\begin{equation}
M_L =  \begin{pmatrix}
0 & 0 &\tilde{y}_1 v_2^3 \\
0 & \tilde{y}_2 v_2^3& \tilde{y}_4 v_2^2 \\
\tilde{y}_3 v_2^3& \tilde{y}_5 v_2^2 & \tilde{y}_6 v_2^1  
\end{pmatrix}.\label{mlmodel}
\end{equation}
The light neutrino mass matrix will have the same structure as this matrix, corresponding to the $A_1$ two-zero texture in the nomenclature of reference \cite{Frampton:2002yf}. This is compatible with the current experiments on neutrino oscillations for the normal neutrino mass ordering and predicts negligible neutrinoless double beta decay effective mass parameter \cite{Ludl:2011vv,Alcaide:2018vni,DeLaVega:2018bkp}. 

Instead of adding extra $SU(2)_L$ doublets $H_2^i$, we can also generate the mass matrix in Eq. (\ref{mlmodel}) by adding flavon fields, $\chi_i$, with a dimension-5 coupling of the type $\tilde{y}^{(\nu)}_i\, \overline{\hat{L}_{i}}\, \tilde{H}\, \hat{S}^c_{i} \chi_i$. The fields $\chi_i$ have $B-L$ and $Z_7$ charges to match those of the $L_i$ and $S_j$ fields.

In this way, we obtained an example of having $M_D$ and $M_R$ diagonal. Notice that in this model, the $U(1)_{B-L}$ can be local, implying the presence of a new $Z^\prime$ gauge boson.

It is possible to turn on a single cLFV process by adding an extra flavon scalar field with charge $\omega^{-2}$ and $+1$ under $B-L$. To have a good phenomenology of neutrino masses and mixings, the charges of the three $SU(2)_L$ $H^{\prime}_{i}$ are $\omega$, $\omega^2$, $\omega^5$. For example, to allow the $\tau \rightarrow e \gamma$, we keep the same charges for the leptons. In this case, we will have the  $m_{\nu}$ matrix has a $B_3$ two-zero texture. To obtain the other cLFV processes, we need to permute the $Z_7$ lepton charges. For the case of $\mu \rightarrow e \gamma$  the $Z_7$ charges of the lepton are: $f_1 \sim \omega$, $f_2 \sim \omega^3$ and $f_3 \sim \omega^2$  with the $B_4$ two-zero texture  in the $m_{\nu}$ matrix. To allow the $\tau \rightarrow \mu \gamma$ the charges of the leptons are: $f_1 \sim \omega^2$, $f_2 \sim \omega^1$ and $f_3 \sim \omega^3$ and a $A_1$ two-zero texture for the $m_{\nu}$ matrix.

Turning the cLFV processes on and off in the inverse seesaw is also possible. Reference ~\cite{Garnica:2023ccx} presents a model for the inverse seesaw based on the $Z_5 \times U(1)_{B-L}$ symmetry that suppresses all the cLFV processes. We will use the same model presented in ~\cite{Garnica:2023ccx} to exemplify how a single cLFV process can be present in the invere seesaw model with an additional flavon scalar field with charges $\omega^{-2}$ and $+1$ under $Z_5$ and $B-L$ respectively. Each case differs in the permutation of the lepton charges and the texture of $m_{\nu}$. For example, in $\tau \rightarrow e \gamma$ process, the order of the charges is the same as the \cite{Garnica:2023ccx}, with $A_1$ two-zero texture for the $m_{\nu}$. For $\mu \rightarrow e \gamma$, we the $Z_5$ lepton charges are: $f_1 \sim \omega$, $f_2 \sim \omega^3$ and $f_3 \sim \omega^2$ this results on $A_2$ two-textures for the $m_{\nu}$ matrix. Finally, to allow  $\tau \rightarrow \mu \gamma$,  the $Z_5$ lepton charges are: $f_1 \sim \omega^2$, $f_2 \sim \omega^1$ and $f_3 \sim \omega^3$, and as a consequence the $m_{\nu}$ has a $B_3$ two-zero texture.

\section{Numerical Analysis}\label{NA}

We present a numerical analysis for the inverse and linear low-scale seesaw realizations, focusing on their allowed parameter space regions. In our approach, we included the non-unitarity limits from neutrino oscillation experiments, as reported in Ref.~\cite{Forero:2021azc} and quoted in Table~\ref{symmetric_constraints}. We have cross-checked our results using two methods for the diagonalization of the neutrino mass matrix $M_\nu$. Specifically, we have employed the well-known block matrix diagonalization method (BMDM) presented in references \cite{Garnica:2023ccx} and an exact numerical diagonalization using the Takagi decomposition implemented in a Phyton routine \footnote{The relevant expressions in the BMDM, for constructing the neutrino mass $M_\nu$ and the relevant matrices used in our analysis are reported in the appendix in Table \ref{BMDM-ls-formulas}.
}. The significance of our findings lies in the detailed understanding of the low-scale seesaw realizations and their parameter space regions.

\begin{table}[ ]
\begin{tabular}{|l|l|}
\hline
Parameters & Constraints  \\ \hline
$\alpha^{\prime}_{11} \leq$ &$3.1\times 10^{-2}$  \\ \hline
 $\alpha^{\prime}_{22}\leq$&$5 \times 10^{-3}$  \\ \hline
 $\alpha^{\prime}_{33} \leq$& $1.1 \times 10^{-1}$ \\ \hline
  $\alpha_{21} \leq$& $1.3 \times 10^{-2}$ \\ \hline
   $\alpha_{31} \leq$& $3.3 \times 10^{-2}$ \\ \hline
    $\alpha_{32} \leq$& $9 \times 10^{-3}$ \\ \hline
\end{tabular}
\caption{\label{symmetric_constraints}Current constraints of the $\alpha$ parameters at 90$\%$ C.L~\cite{Forero:2021azc}. The $\alpha^{\prime}_{ii}=1-\alpha_{ii}$.}
\end{table}

We are interested in low-scale seesaw scenarios where the cLFV effects may be strongly suppressed due to a particular structure of the $\eta$ matrix that characterizes the non-unitary effects due to the presence of the new NHL.  Because these non-unitarity limits are extracted from the symmetric parametrization \cite{Schechter:1981cv,Rodejohann:2011vc}, we need an equivalence between the symmetric and the $\eta$ parametrizations \cite{Blennow:2016jkn, Celestino-Ramirez:2024gmq}, which is given by: 
\begin{align}
\label{matching}
\eta&= \frac{1}{2}\begin{pmatrix}
2\alpha_{11} & \alpha^{*}_{21} & \alpha^{*}_{31} \\ 
\alpha_{21} & 2\alpha_{22} &\alpha^{*}_{32} \\
\alpha_{31} &  \alpha_{32} & 2\alpha_{33}
\end{pmatrix}.
\end{align}


Let us consider the inverse and linear seesaw models defined by the neutrino mass matrices in Eqs. (\ref{inverse seesaw}) and (\ref{Linear seesaw}); respectively, and we consider the case when $M_D$ and $M$ are diagonal matrices, given by 
\begin{align}
\label{M}
    M_{3\times 3} &=   \textrm{v}_M \cdot diag(1+\epsilon_{M_{11}},1+\epsilon_{M_{22}},1+\epsilon_{M_{22}}), \\ 
\label{M_D}    M_{D_{3\times 3}} &= \frac{v_{SM}}{\sqrt{2}} \cdot diag(Y_{11},Y_{22},Y_{33}),
\end{align}
with $v_{\textrm{SM}}=246$ GeV the vacuum expectation value (vev) of the SM Higgs field, whereas $  \textrm{v}_M$ is the mass scale associated with the new heavy states. 

For the \textbf{inverse seesaw case}, the values of the $\mu$  matrix are obtained from Eq. (\ref{mass_inverseseesaw}), which relates it to the physical light neutrino masses and neutrino oscillation parameters. Therefore, the neutrino mass matrix in Eq. (\ref{inverse seesaw}) involves 13 parameters: the three mixing angles ($\theta_{12}$,$\theta_{13}$,$\theta_{23}$), a Dirac CP-phase, the three physical light neutrino mass, $m^{diag}_{\nu}=(m_{\nu_{1}},m_{\nu_{2}},m_{\nu_{3}})$, the three Yukawa couplings  ($Y_{11}$, $Y_{22}$, $Y_{33}$) for the $M_D$ matrix and another three parameters ($\epsilon_{M_{11}}$, $\epsilon_{M_{22}}$, $\epsilon_{M_{33}}$) from the $M$ matrix.

On the other hand, regarding the \textbf{linear seesaw model}, the matrix $M_L$ can be written in terms of the $\eta$ and $m_\nu$ matrices as
 \begin{equation}
\label{M_L}
 M_L= \begin{pmatrix}
\frac{1}{\sqrt{2}}\sqrt{|\eta^{-1}|}_{11}m_{\nu_{11}} & \sqrt{2|\eta^{-1}|}_{22}m_{\nu_{12}}\cdot x_1 & \sqrt{2|\eta^{-1}|}_{33}m_{\nu_{13}}\cdot x_2\\ 
\sqrt{2|\eta^{-1}|}_{11}m_{\nu_{12}}\cdot (1-x_1) & \frac{1}{\sqrt{2}}\sqrt{|\eta^{-1}|}_{22}m_{\nu_{22}} &\sqrt{2|\eta^{-1}|}_{33}  m_{\nu_{23}}\cdot x_3\\
\sqrt{2|\eta^{-1}|}_{11}m_{\nu_{13}}\cdot (1-x_2) &  \sqrt{2|\eta^{-1}|}_{22}m_{\nu_{23}}\cdot (1-x_3) & \frac{1}{\sqrt{2}}\sqrt{|\eta^{-1}|}_{33}m_{\nu_{33}}
\end{pmatrix}, 
\end{equation}
where the parameters $x_1$, $x_2$, and $x_3$ are real free parameters. Thus, the neutrino matrix in Eq. (\ref{inverse seesaw}) is described in terms of the same 13 parameters (used in the inverse case) plus three parameters that represent a percentage of the component of the $m_{\nu}$ matrix ($x_1$, $x_2$, $x_3$) in Eq. (\ref{M_L}). 
We randomly scan the above parameters and fix the scale $v_{SM}=246$ GeV. On the other hand, we take values for $  \textrm{v}_M$ in the range [$10^{-1}-10^2$] TeV logarithmically distributed. Here, we list the random parameters in our analysis:
\begin{itemize}
    \item The neutrino oscillation parameters (mixing angle, squared mass differences and CP-phase) are varied at 3$\sigma$ range using the current oscillation neutrino data in the Table \ref{oscillation parameters}, while the sum of the neutrino masses is kept below the cosmological bound~\cite{deSalas:2020pgw}.
    \item The $\epsilon_{M_{ii}}$ of the $M$ matrix are varied in the region [-0.5,0.5].
    \item The $Y_{ii}$ of the $M_D$ matrix are varied in the region [-0.5,0.5].
    \item The parameters $x_i$ in the $M_L$ matrix are varied in the region [0,1].
\end{itemize}

\begin{table}[h!]
\def\arraystretch{2}
\begin{tabular}{|c|c|}
\hline
Parameters & Normal hierarchy at 3$\sigma$ \\ \hline
       $\Delta m^2_{21}$($eV^2$)    & $(6.94-8.14)\times 10^{-5}$                      \\ \hline
$\Delta m^2_{31}$($eV^2$)           & $(2.47-2.63)\times 10^{-3}$                      \\ \hline
$\theta_{12}/^{\circ}$           &   31.4-37.4                    \\ \hline
   $\theta_{23}/^{\circ}$        & 41.2-51.33                      \\ \hline
$\theta_{13}/^{\circ}$           & 8.13-8.92                      \\ \hline
    $\delta/^{\circ}$       &   128-359                    \\ \hline
\end{tabular}
\caption{\label{oscillation parameters}Numerical inputs for the oscillation parameters used in our analysis \cite{deSalas:2020pgw}.}
\end{table}
Then, we keep in our analysis the mass hierarchy between the $M_D$ and $M$ matrices. We can diagonalize the inverse's and linear seesaw's full mass matrix to obtain the full mixing matrix. The computation of the non-unitary matrix is an important difference between the BMDM and the full diagonalization process. In the case of BMDM, we compute the non-unitary matrix using the Eq. (\ref{non_unitary_effect-O}). On the other hand, we will extract the non-unitary matrix in the full diagonalization process using the whole neutrino mixing matrix. From Eq. (\ref{KL}), we can obtain the non-unitary matrix as follows:
\begin{equation}
    \eta=\frac{1}{2}(\mathbb{I}_{3 \times 3}-K_{L_{3\times3}}K^{\dagger}_{L_{3\times3}}),
\end{equation}
where $K_L$ is the light submatrix of the whole neutrino mixing matrix. We finally put conditions to fulfill the mass hierarchy. Now, we can compute the non-unitary effects in the different scenarios with all these elements.

\section{Results}

Now that we have explained our numerical procedure, we can show and discuss our results, which are presented considering two cases, A and B. Case A considers the $\eta$ matrix diagonal leading to strongly suppressed rates for all the three $\mu-e$, $\tau-e$, and $\tau-\mu$ cLFV possible channels. In contrast, Case B (this is done only for the inverse seesaw) considers some non-zero off-diagonal elements for the $\eta$ matrix allowing sizable rates for a single cLFV channel. 

\begin{figure}[ht]
\includegraphics[scale=0.5]{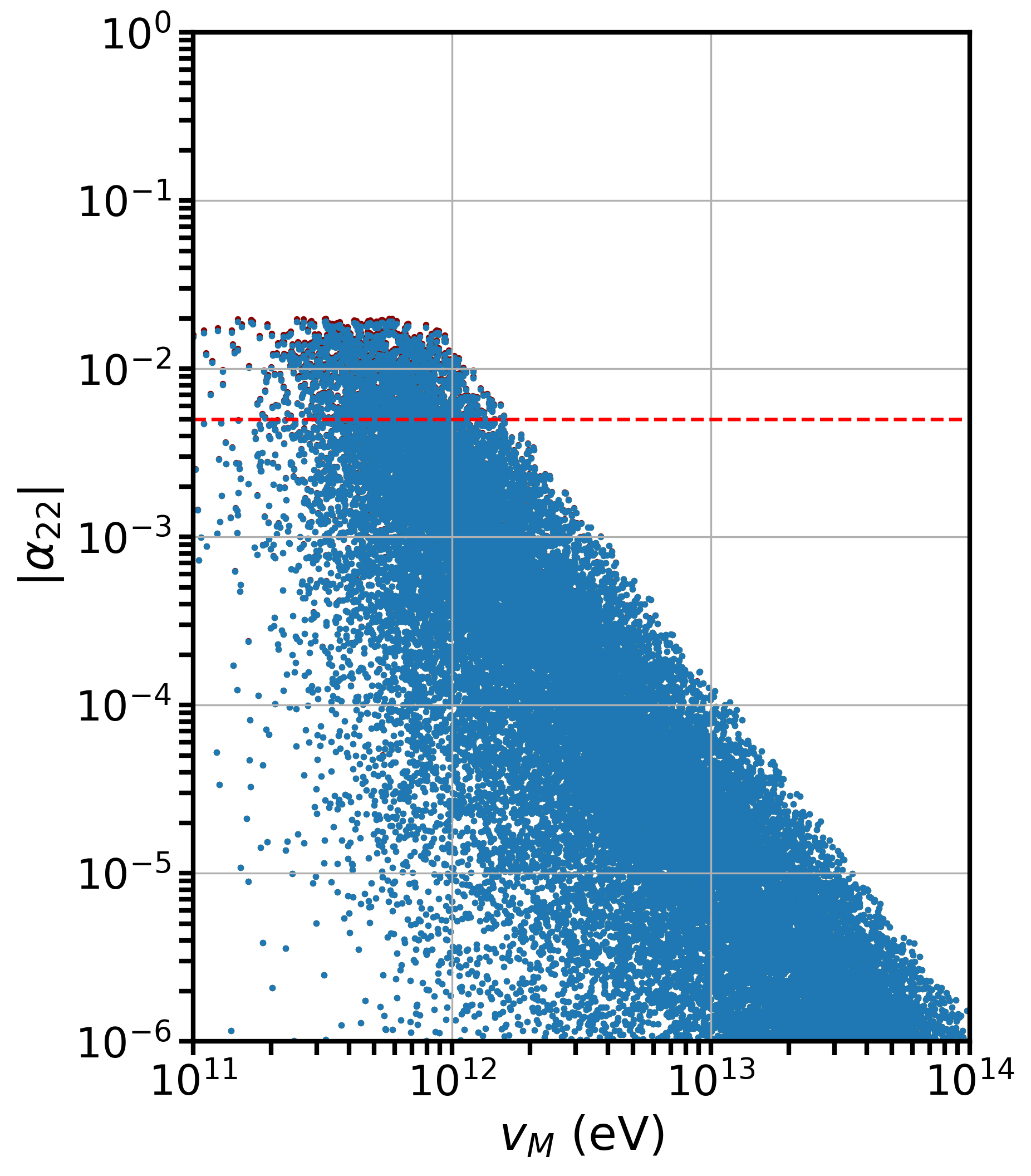}
\label{fig:a_22_constraints}
\caption{The $  \textrm{v}_M$ scale associated with the mass of NHL in low-scale seesaw models vs the non-unitary effects considering the case A. The points in our scan consistent with neutrino oscillation data using the BMDM (full diagonalization method) are represented in blue color (red). The red dashed line corresponds to the limit on the most restricted non-unitary parameter $\alpha_{22}$. We have not encountered significant differences between inverse and linear models for this plot. }
\label{fig:figure1}
\end{figure}

For the \textbf{Case A} the cLFV processes are suppressed. In Fig.~\ref{fig:figure1} we show the $  \textrm{v}_M$ sensitivity using the most restricted non-unitary parameter ($\alpha_{22}$) on the oscillation experiments. It is important to remark, that each case has the same inputs and, as we can see, exists an overlap effect in the plot. We can notice from this figure that, for values $  \textrm{v}_M>2$~TeV, all the points in our analysis are allowed, making this scale a more natural choice for the heavy sector.  

For the \textbf{Case B:} we consider the following textures for the $M$ matrix:  
\eq{
M&=v_{M}\begin{pmatrix}
1+\epsilon_{11} & 1+\epsilon_{21} & 0\\ 
0& 1+\epsilon_{22} & 0\\
0 & 0  & 1+\epsilon_{33}
\end{pmatrix}, \label{e-mu}\\ 
\nonumber \\
M&=  \textrm{v}_M\begin{pmatrix}
1+\epsilon_{11} & 0 & 1+\epsilon_{31}\\ 
0& 1+\epsilon_{22} & 0\\
0 & 0  & 1+\epsilon_{33}
\end{pmatrix}, \label{e-tau}\\
\nonumber \\
M&=  \textrm{v}_M\begin{pmatrix}
1+\epsilon_{11} & 0 & 0\\ 
0& 1+\epsilon_{22} & 1+\epsilon_{32} \\
0 & 0 & 1+\epsilon_{33}
\end{pmatrix}.\label{mu-tau}
}
The matrices in Eqs. (\ref{e-mu}), (\ref{e-tau}), and (\ref{mu-tau}) generates $\mu-e$, $\tau-e$, and $\tau-\mu$ cLFV transitions, respectively. Note that in this case, additionally to the diagonal $\epsilon_{M_{ii}}$ parameters, we have here to consider the ($\epsilon_{21}$, $\epsilon_{31}$, or $\epsilon_{21}$) for each particular case, with these parameters varying in the same region, which is in the ([-0.5,0.5]) range.

\begin{figure}[ht]
    \begin{subfigure}{0.3\textwidth}
    \includegraphics[width=\textwidth]{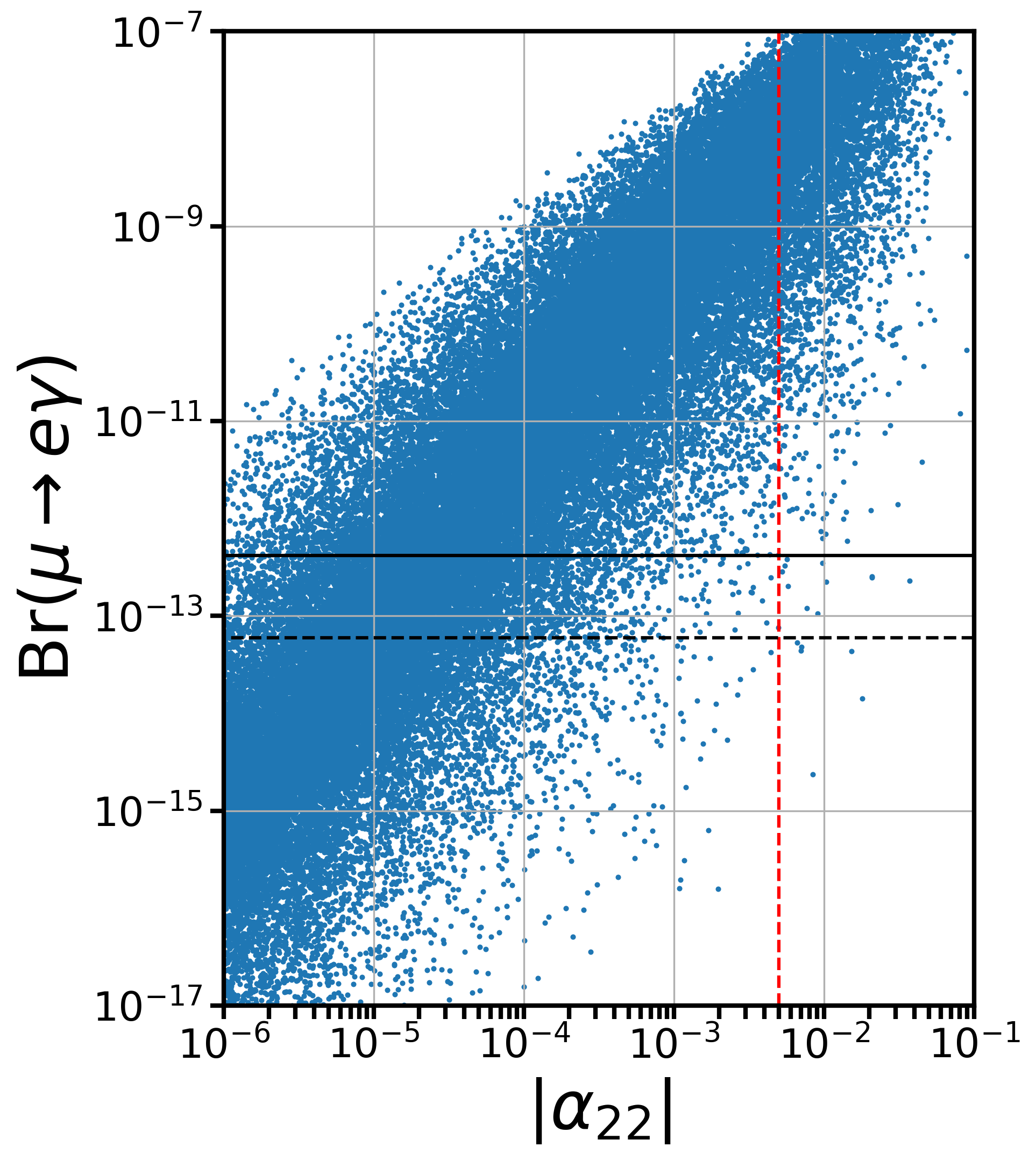}
    \end{subfigure} 
    \begin{subfigure}{0.3\textwidth}
    \includegraphics[width=\textwidth]{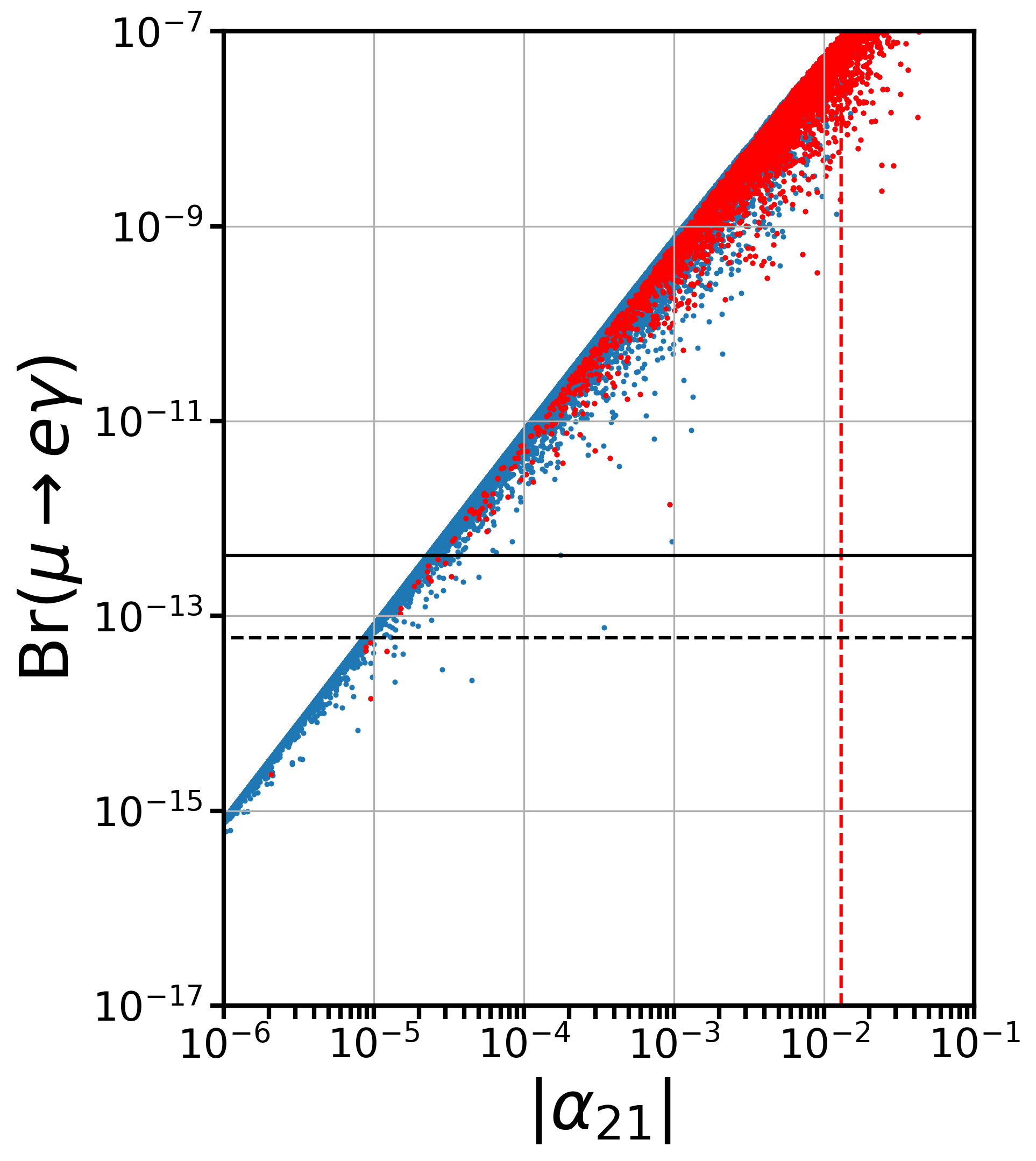}
    \end{subfigure}
    \begin{subfigure}{0.3\textwidth}
    \includegraphics[width=\textwidth]{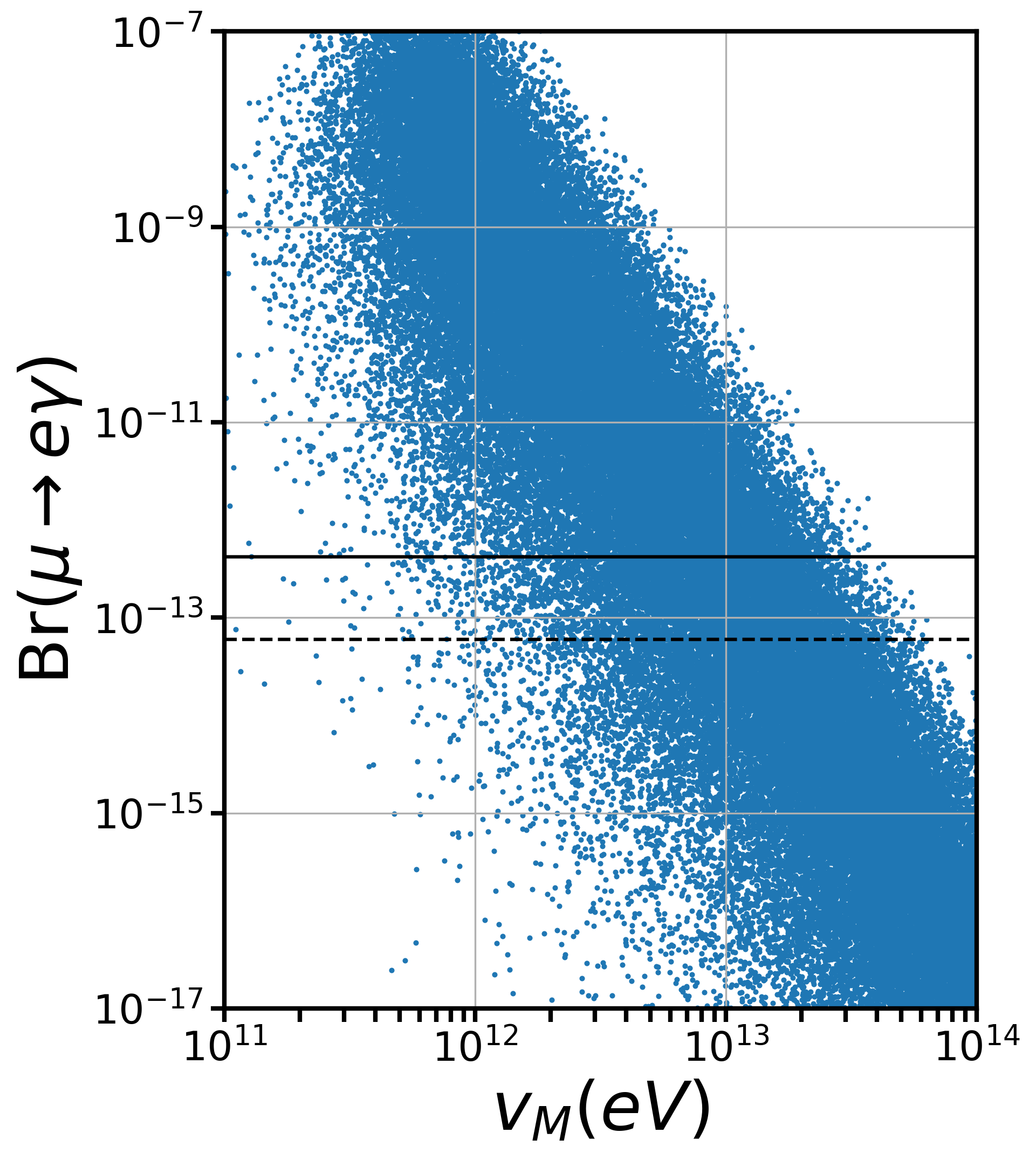}
    \end{subfigure}
    \begin{subfigure}{0.3\textwidth}
    \includegraphics[width=\textwidth]{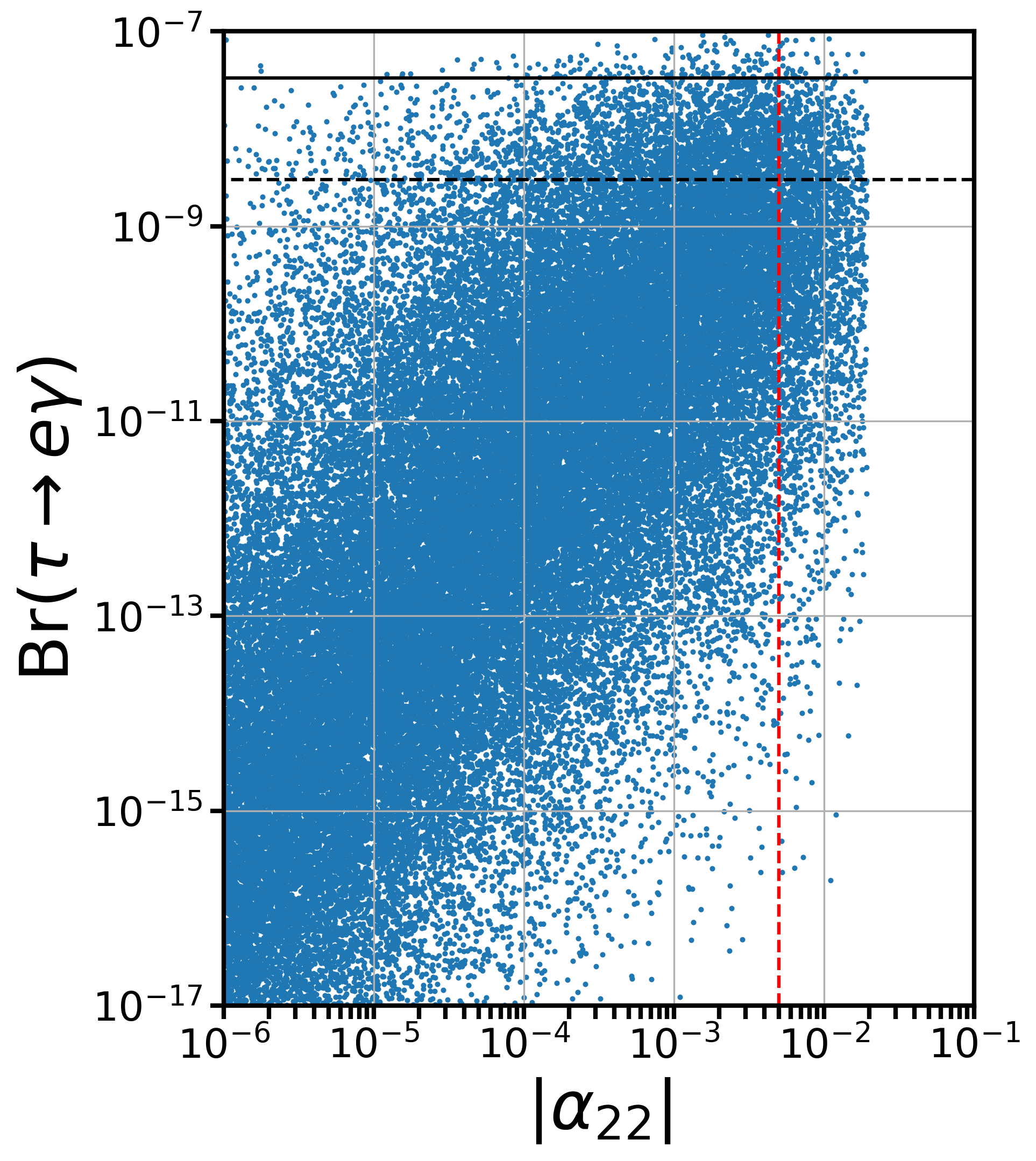}
    \end{subfigure}
    \begin{subfigure}{0.3\textwidth}
    \includegraphics[width=\textwidth]{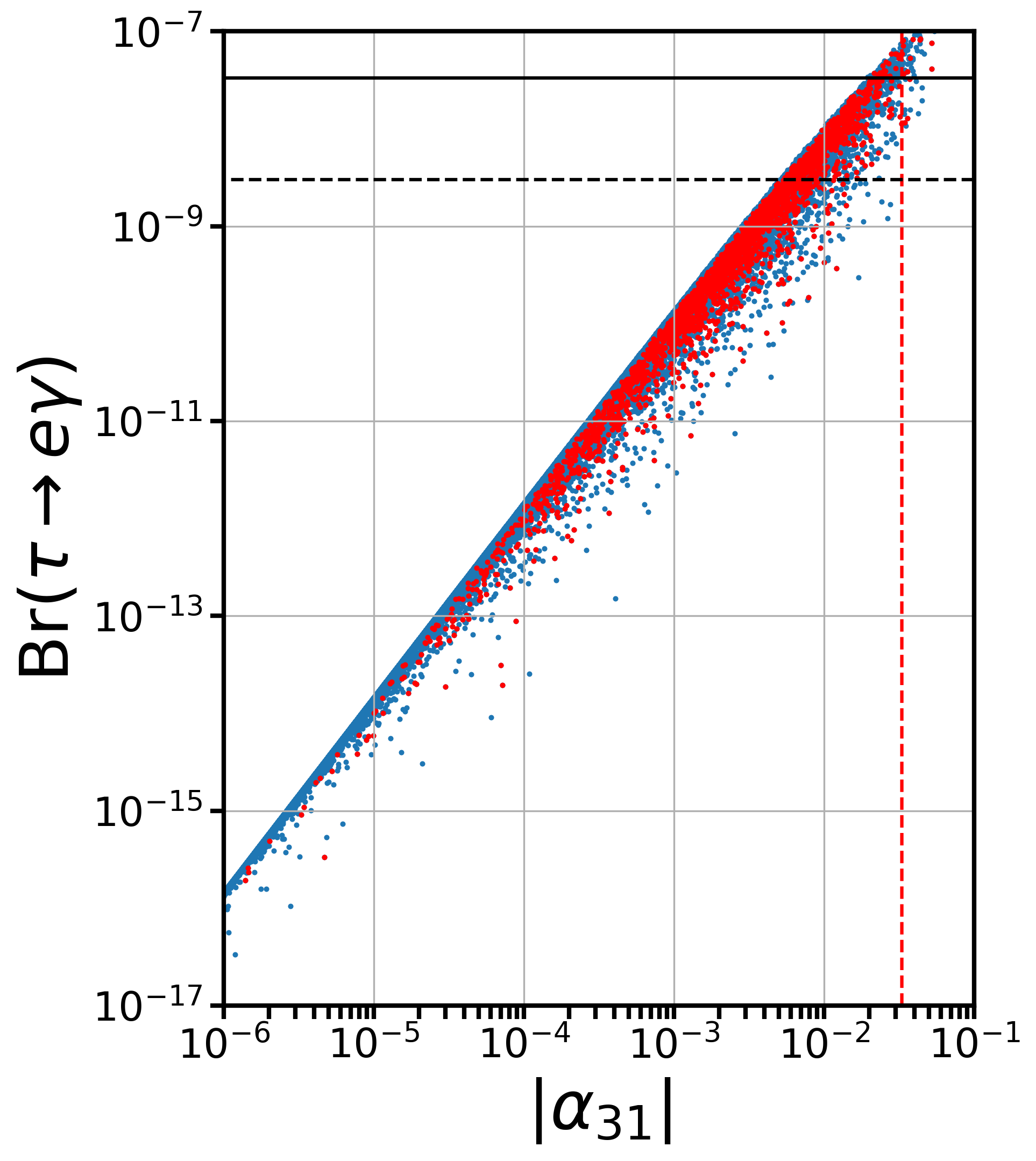}
    \end{subfigure}
    \begin{subfigure}{0.3\textwidth}
    \includegraphics[width=\textwidth]{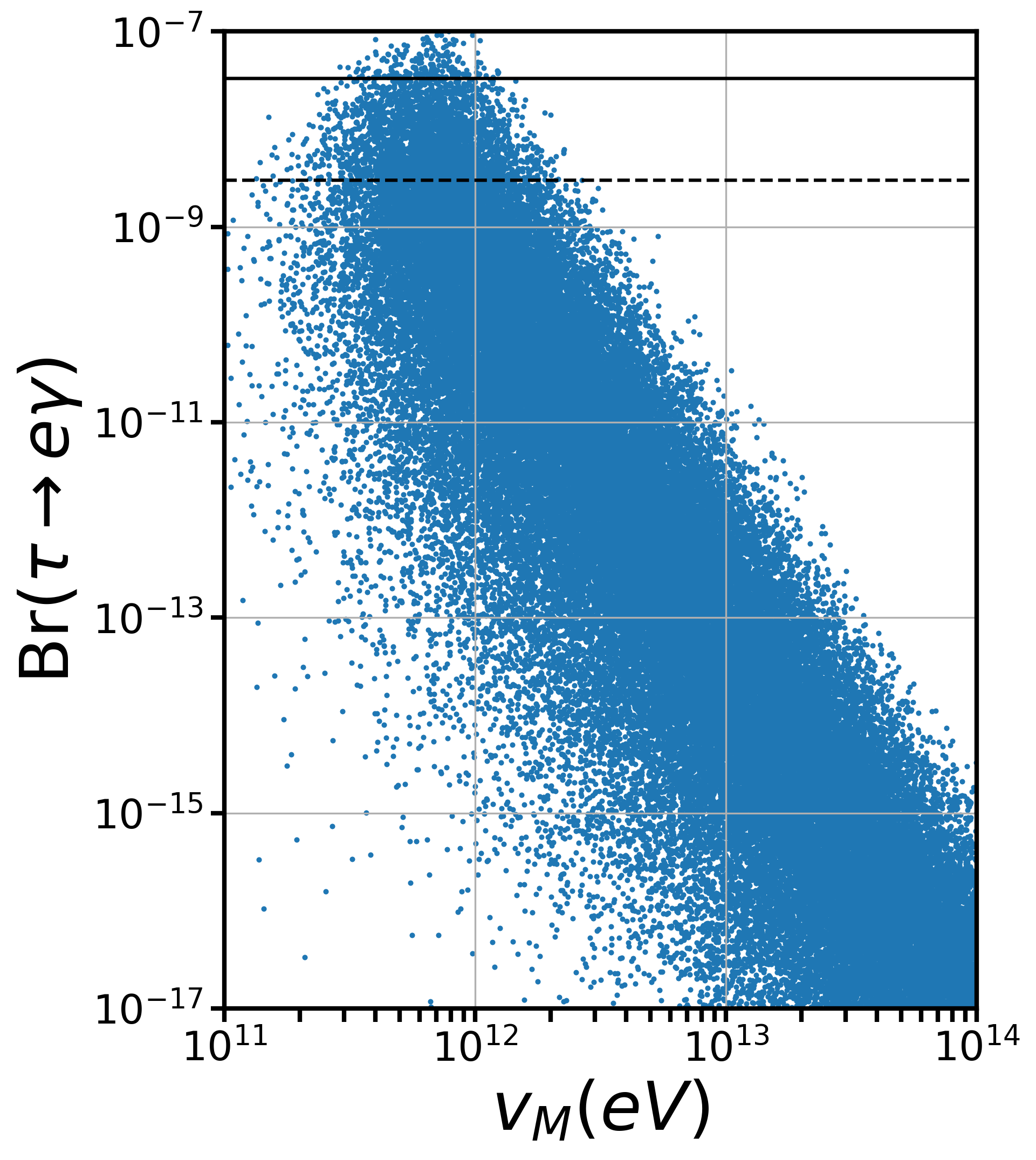}
    \end{subfigure}
     \begin{subfigure}{0.3\textwidth}
    \includegraphics[width=\textwidth]{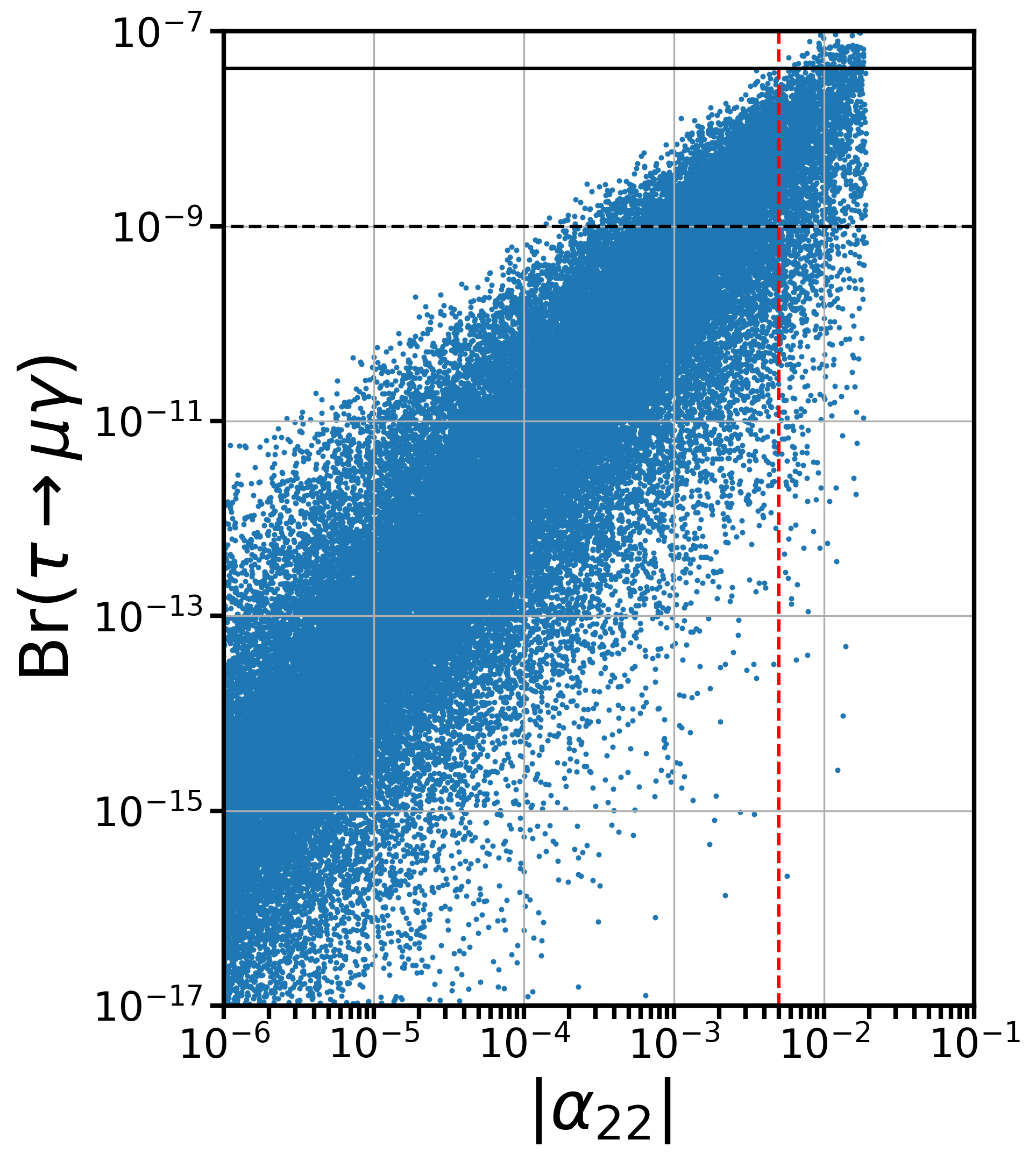}
    \end{subfigure}
    \begin{subfigure}{0.3\textwidth}
    \includegraphics[width=\textwidth]{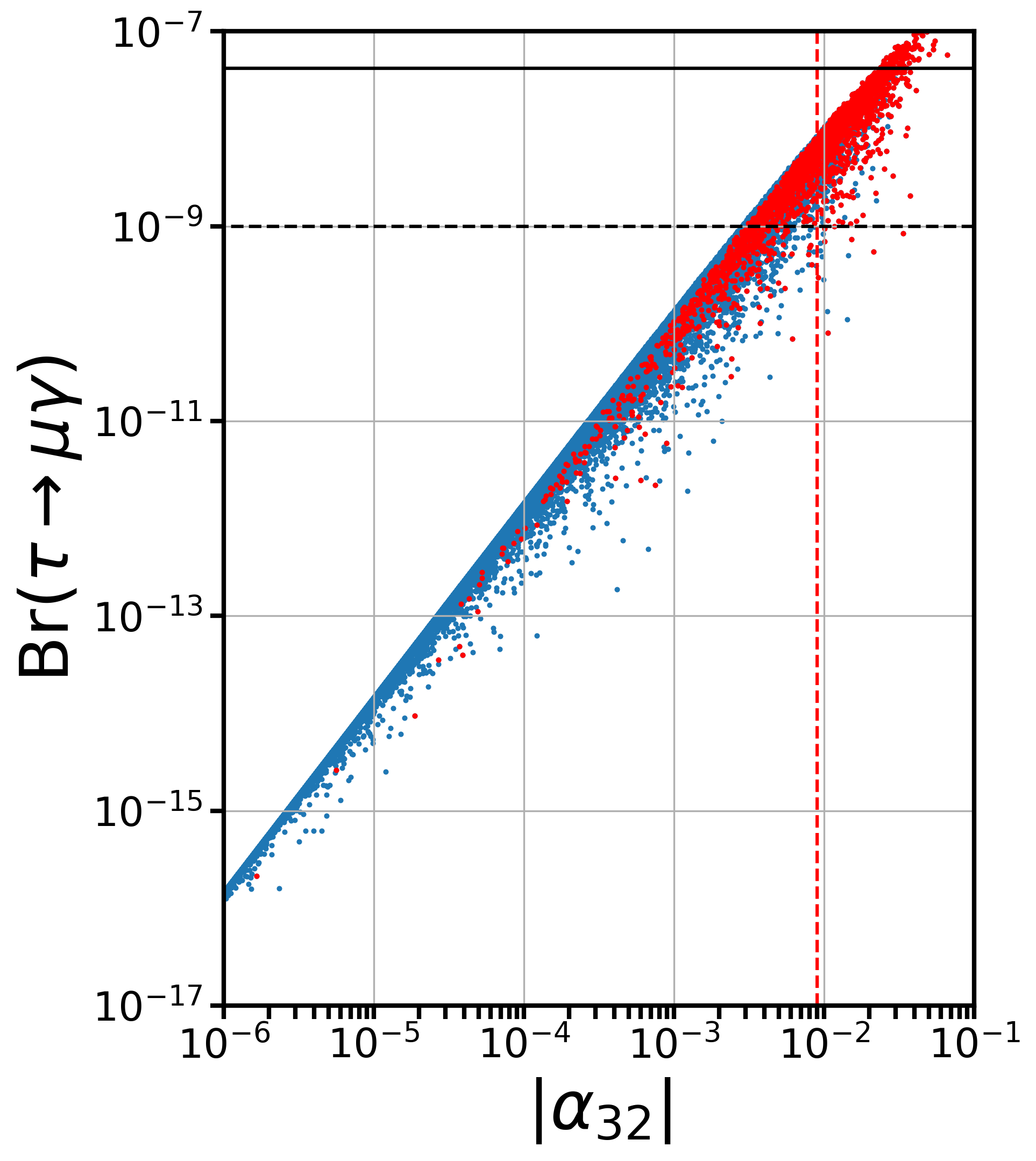}
    \end{subfigure}
      \begin{subfigure}{0.3\textwidth}
    \includegraphics[width=\textwidth]{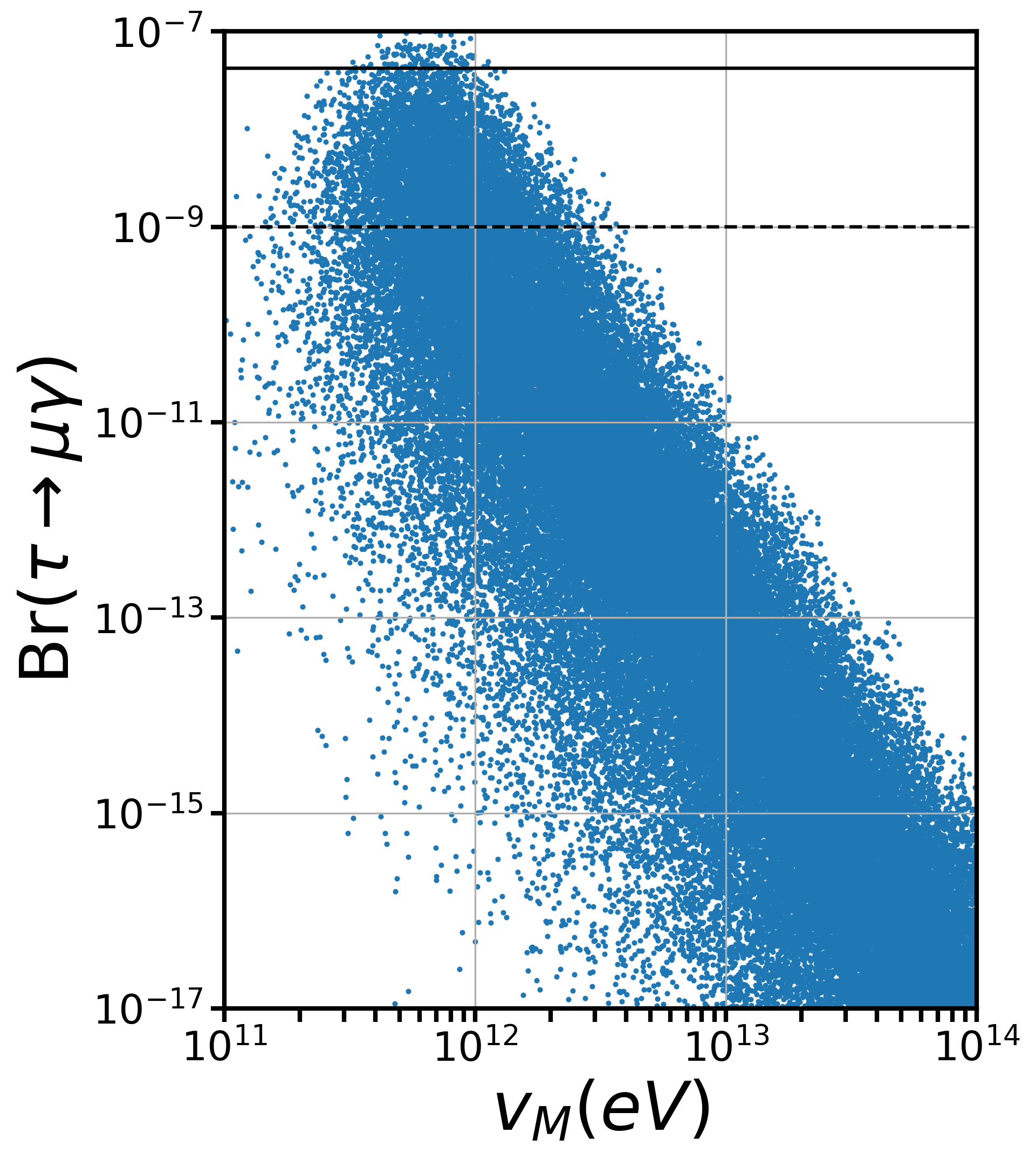}
    \end{subfigure}
\caption{\label{muon_to_electron}We show the parameter space of the Br$(l_{i}\rightarrow l_{j} \gamma )$ in terms of the non-unitary parameters and the $  \textrm{v}_M$ for each case. The black dash (solid) is the process's future (current) limit. Meanwhile, the red dash vertical line is the current constraint of the non-unitary parameters. The red points represent the excluded parameter space using the $\alpha_{22}$ constraint and, simultaneously, allowed by the current and future limit of the process. }
\end{figure}

\begin{figure}
\begin{subfigure}{0.3\textwidth}
    \includegraphics[width=\textwidth]{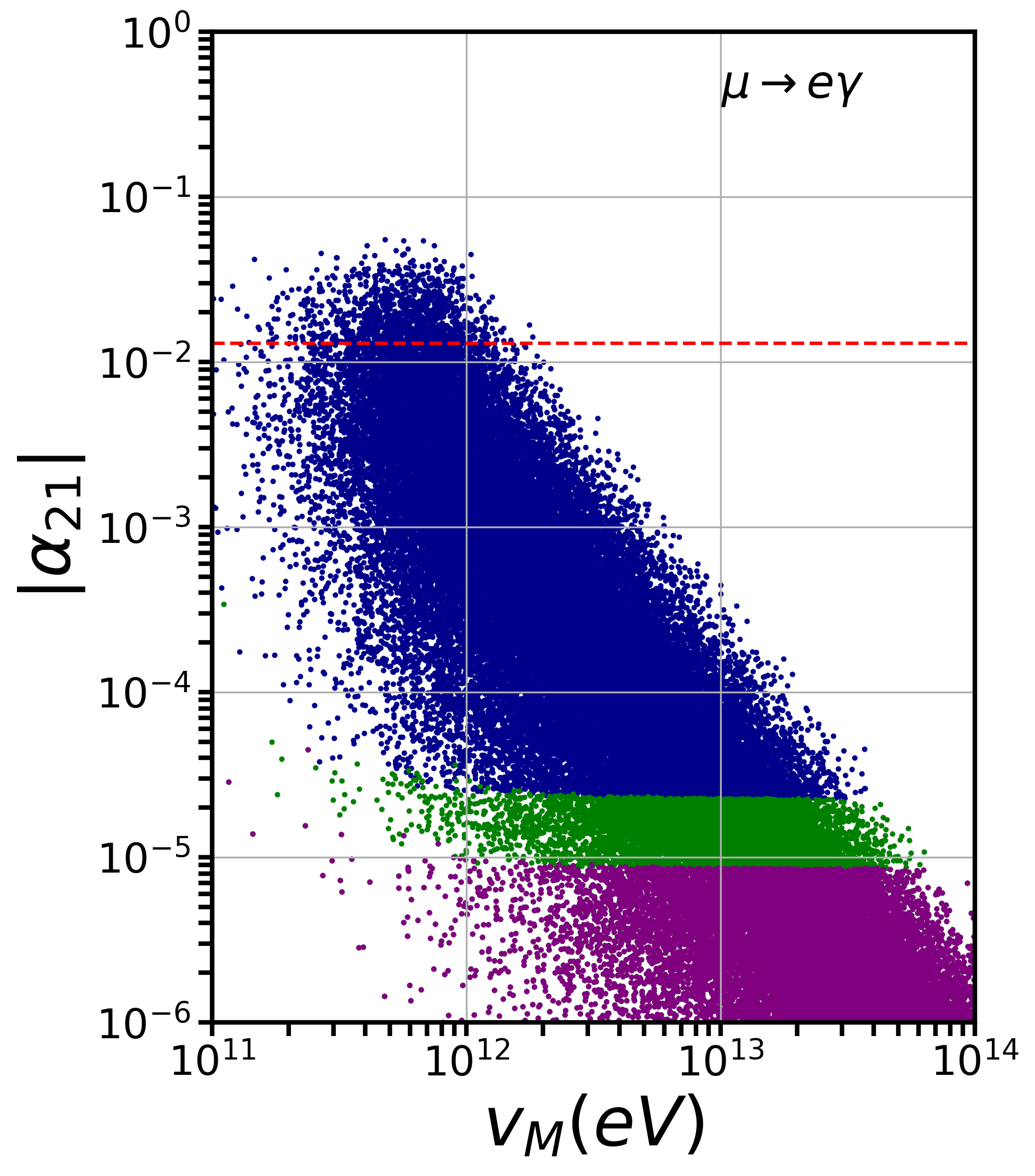}
    \end{subfigure}
\begin{subfigure}{0.3\textwidth}
    \includegraphics[width=\textwidth]{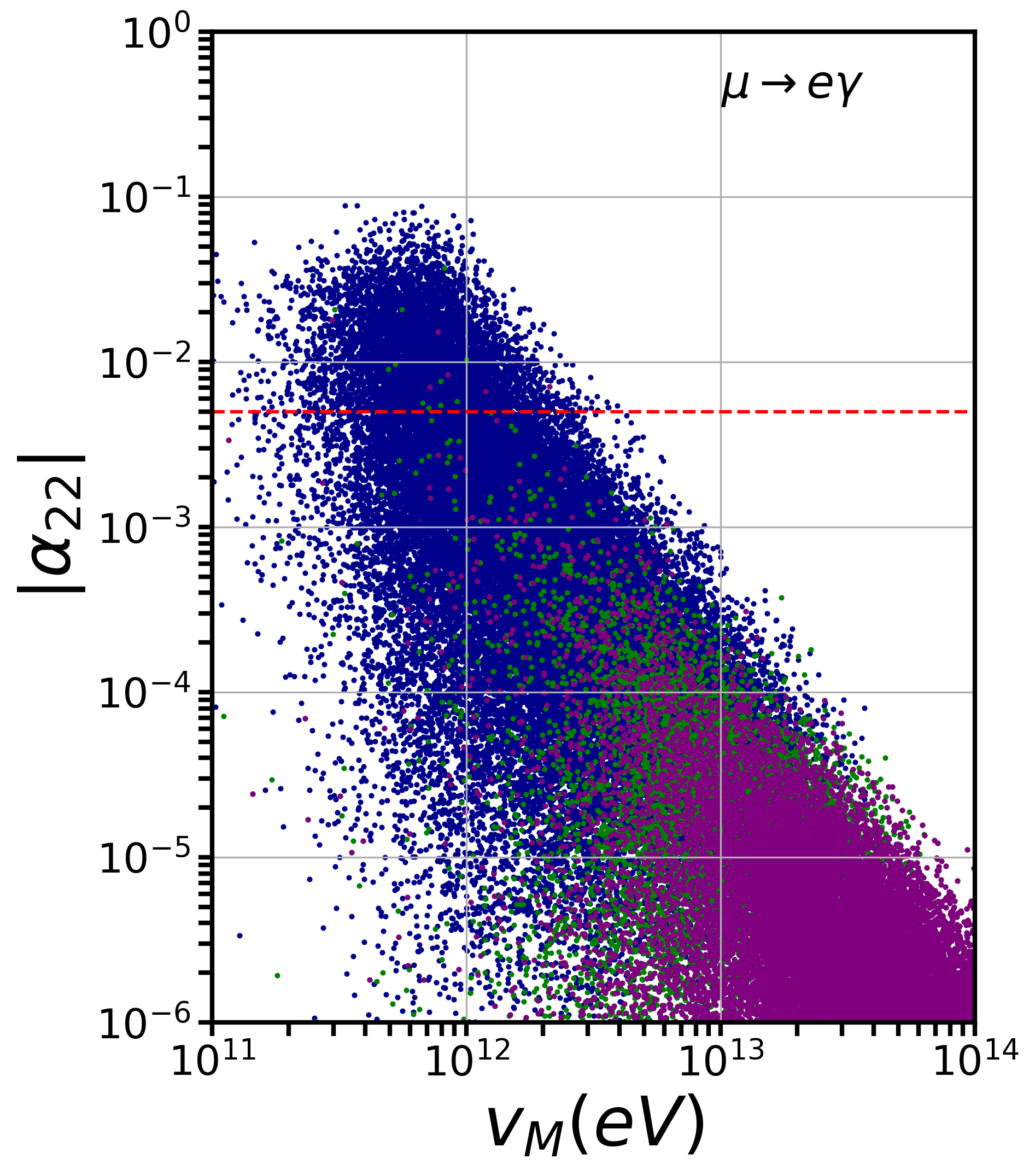}
    \end{subfigure}
 \begin{subfigure}{0.3\textwidth}
    \includegraphics[width=\textwidth]{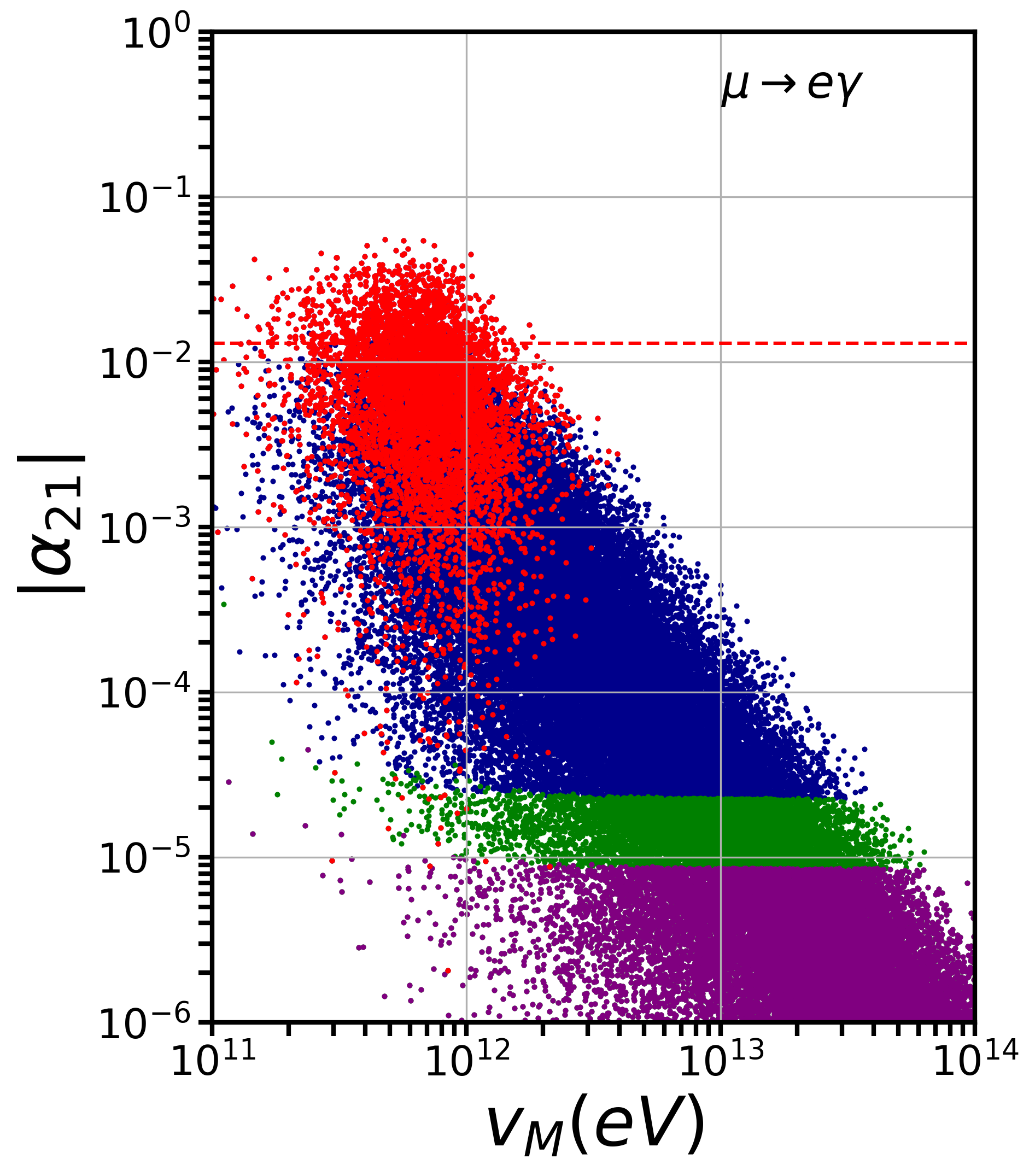}
    \end{subfigure}
    \begin{subfigure}{0.3\textwidth}
    \includegraphics[width=\textwidth]{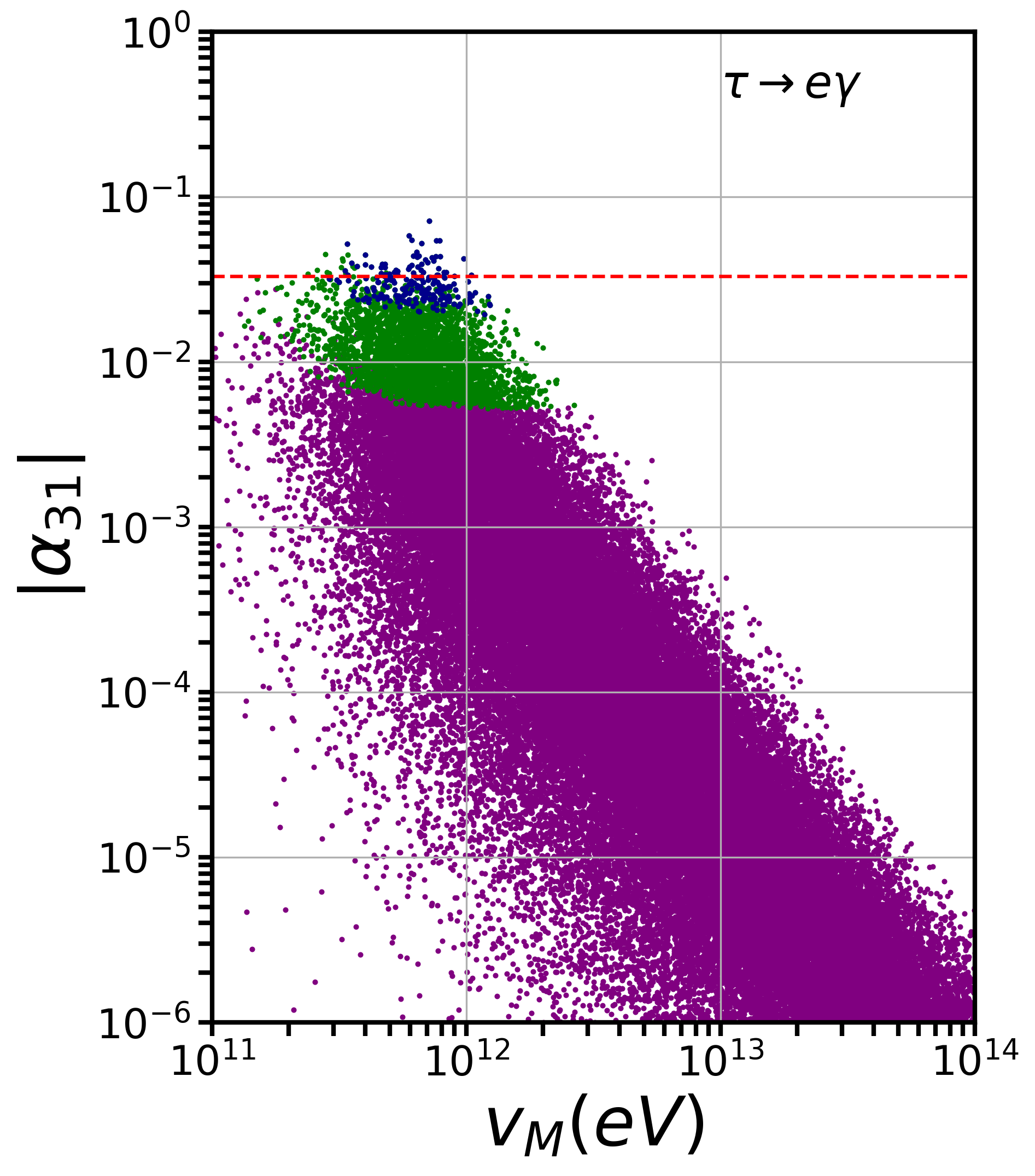}
    \end{subfigure}
    \begin{subfigure}{0.3\textwidth}
    \includegraphics[width=\textwidth]{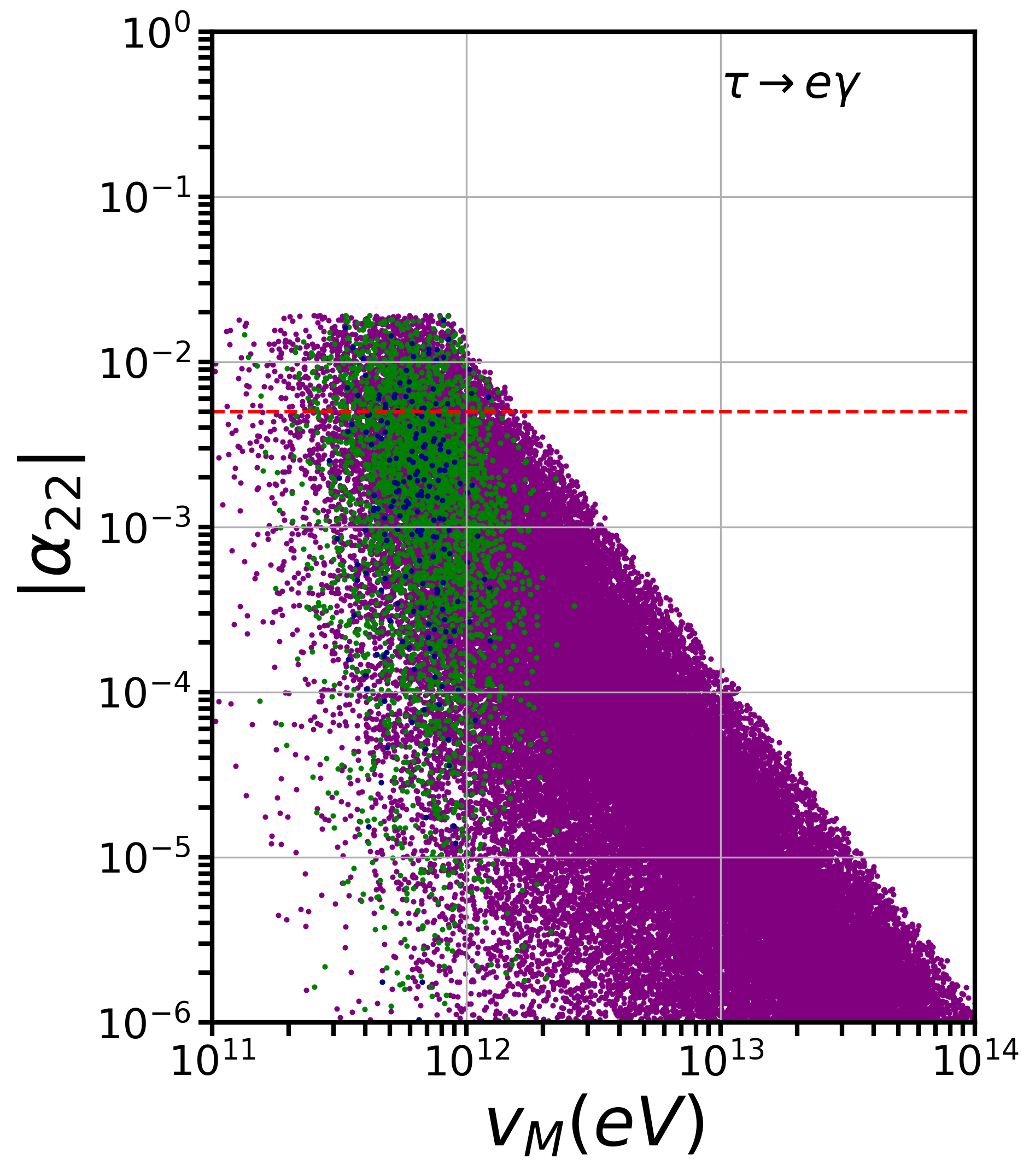}
    \end{subfigure}
\begin{subfigure}{0.3\textwidth}
    \includegraphics[width=\textwidth]{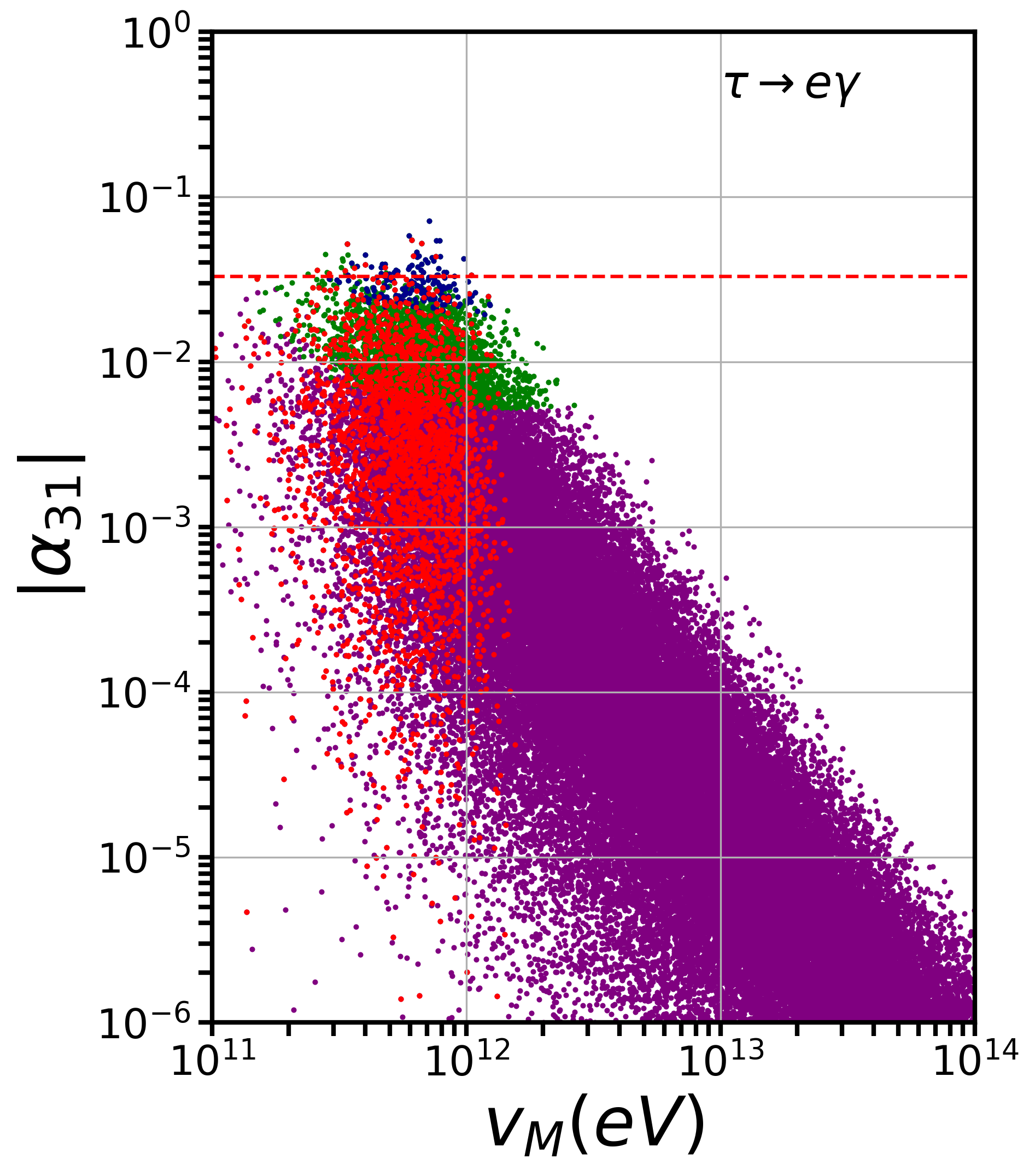}
    \end{subfigure}

    \begin{subfigure}{0.3\textwidth}
    \includegraphics[width=\textwidth]{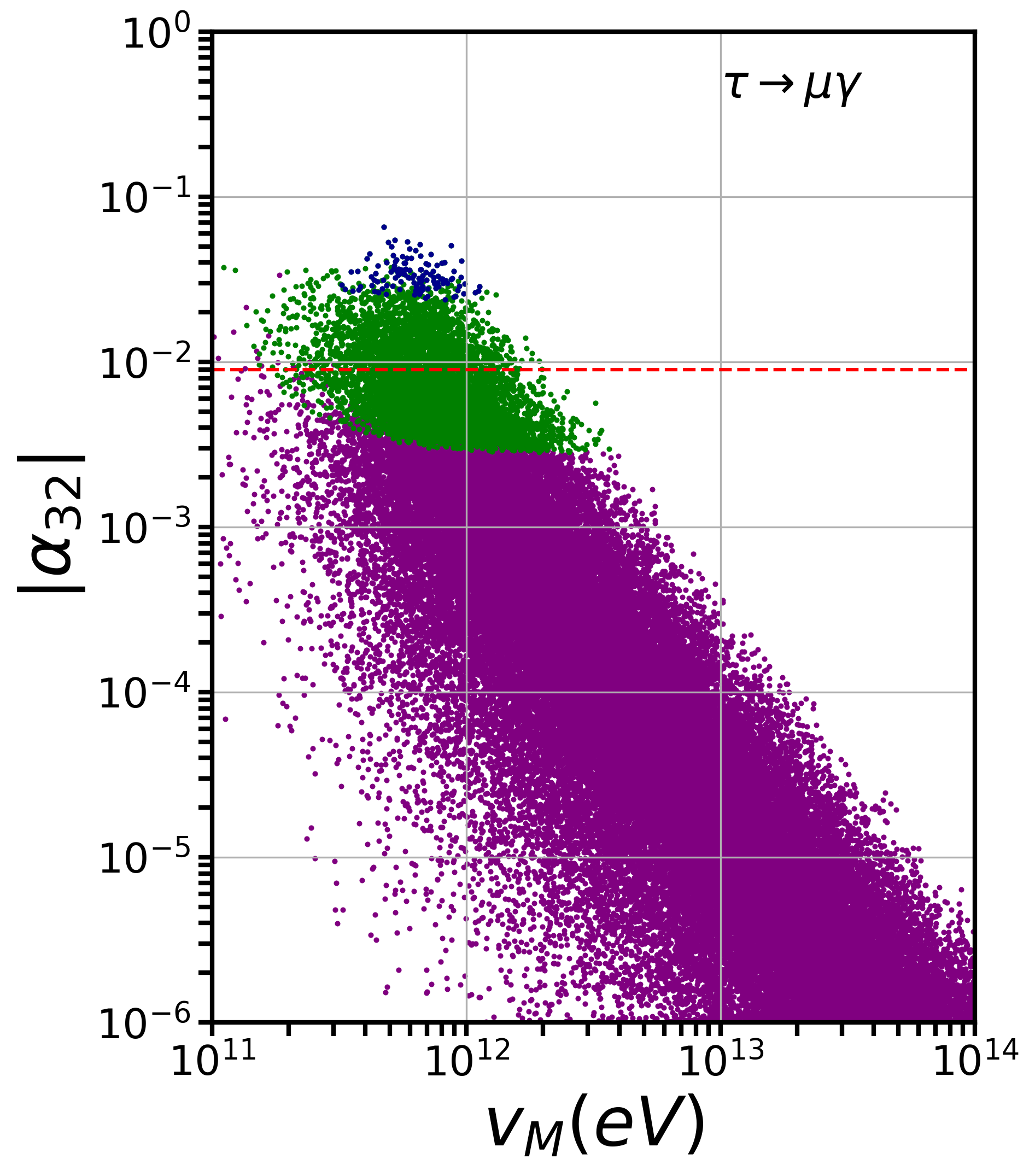}
    \end{subfigure}
\begin{subfigure}{0.3\textwidth}
    \includegraphics[width=\textwidth]{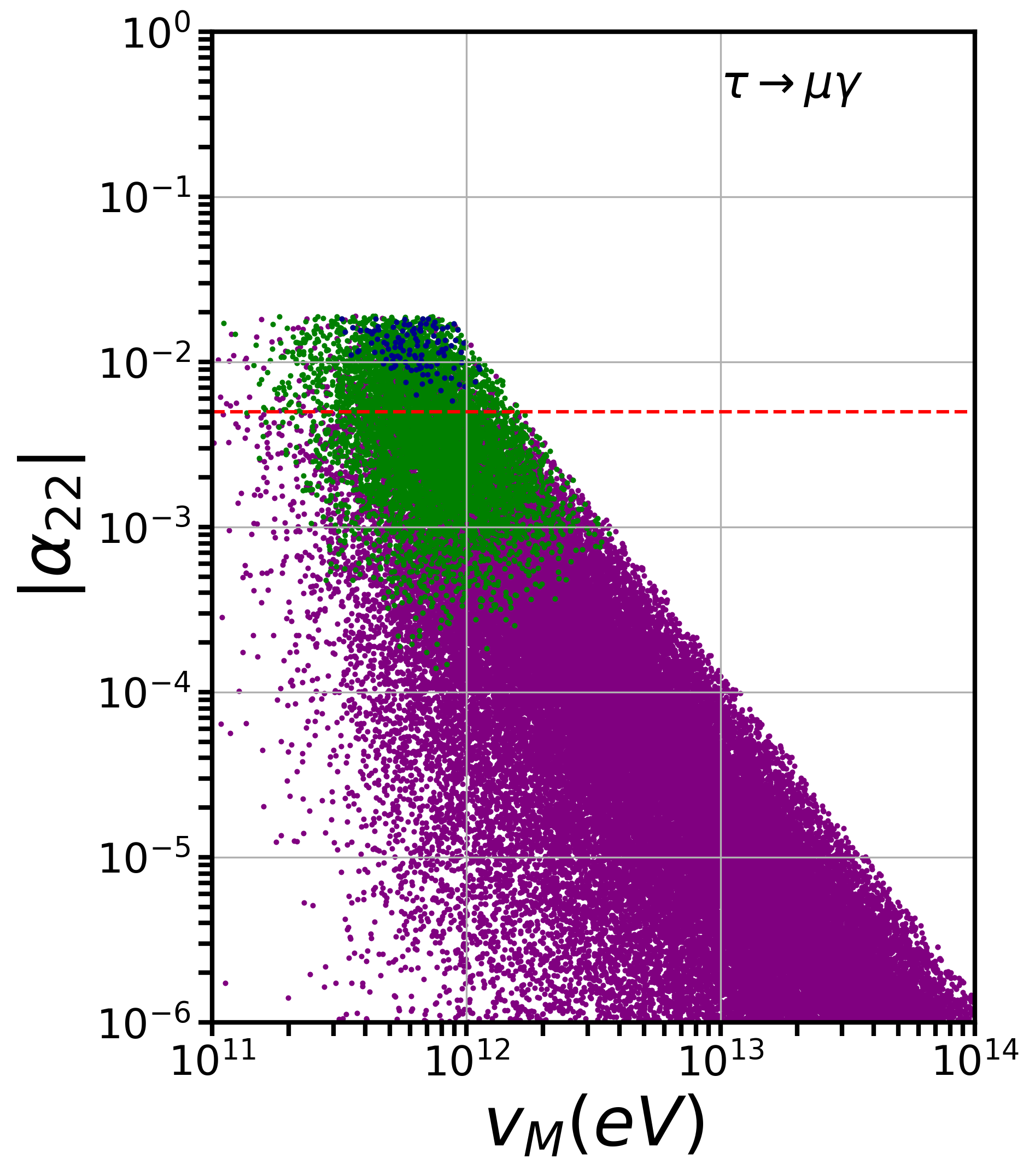}
    \end{subfigure}
  \begin{subfigure}{0.3\textwidth}
    \includegraphics[width=\textwidth]{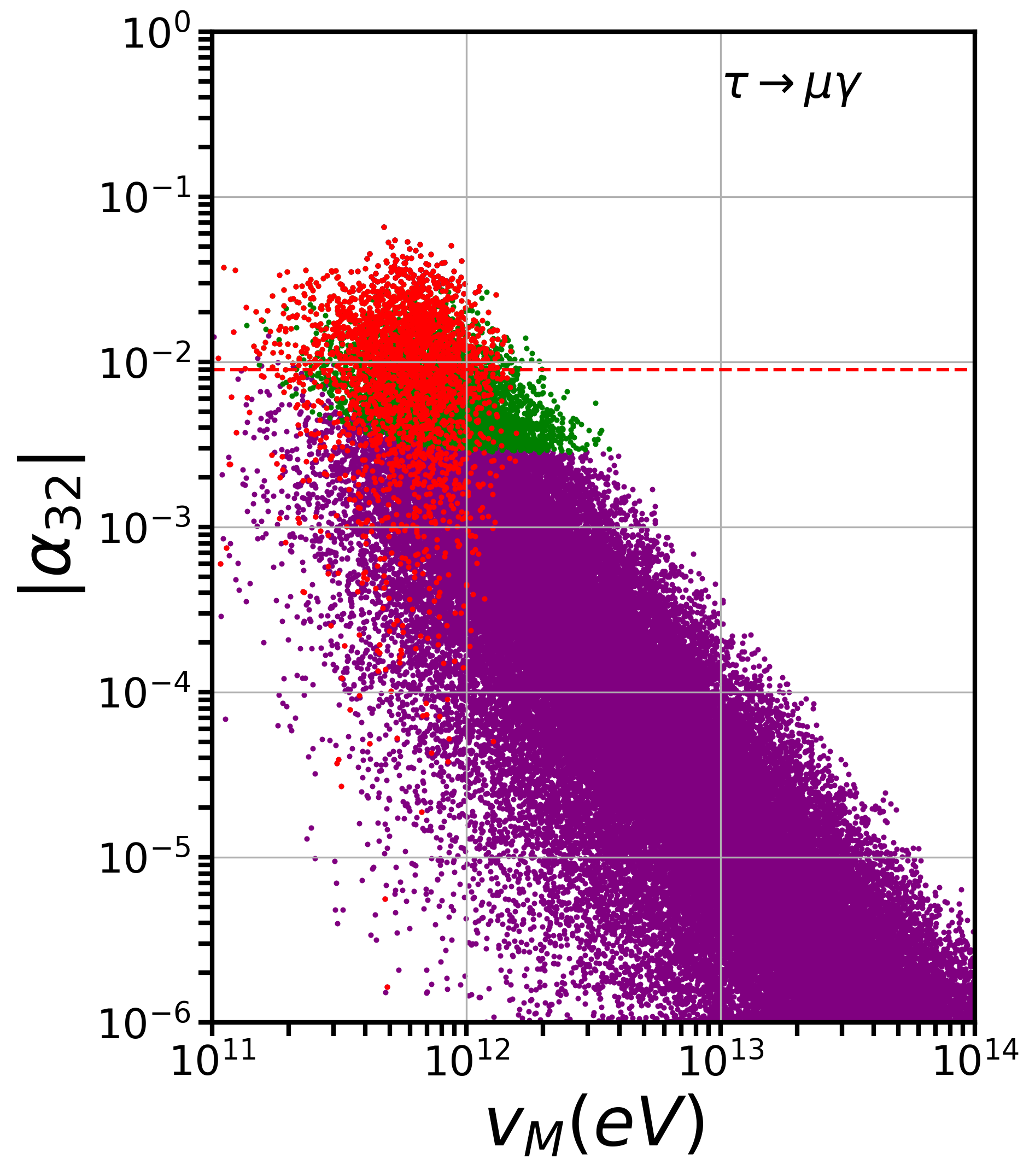}
    \end{subfigure}
\caption{\label{tau_to_electron}
We plot the parameter space of the cLFV processes in the $|\alpha_{ij|}$-$  \textrm{v}_M$ planes. The red dashed line is the current limit of the non-unitary parameters. Each row represents a different cLFV process. The dark blue points represent the parameter space excluded by the current cLFV limits,  the green points are the additional excluded parameter space when including the future limit case, and the purple points are the ones allowed by future cLFV limits. The sum of these dark blue, green, and purple points represent the same parameter space as the blue points in the previous figure. The red points are the parameter space excluded by the $\alpha_{22}$ limit.}
\end{figure}

In the plots presented in Figs.~\ref{muon_to_electron} and \ref{tau_to_electron}, we show an analysis of the non-unitary parameters using their current oscillation constraints (shown in Table~\ref{oscillation parameters}) as well as the current and future limits on the cLFV branching ratios (Table~\ref{table cLFV limits}). 
We performed an exhaustive scan and show here different cases for the parameter space of the branching ratios and the non-unitary parameters (or $\textrm{v}_{M}$). The constraints from different experimental results for the cLFV branching ratios and for non-unitary are also shown in the figures. From the panels in Fig.~\ref{muon_to_electron} we notice that the $\alpha_{22}$ constraint is more efficient in restricting a region of parameter space that is still allowed by the cLFV restrictions, showing the complementarity of both limits. Other non-unitary parameters are clearly less effective, except for the case of $\alpha_{32}$.

The first row of Fig.~\ref{muon_to_electron} also shows that, as expected, the non-unitary constrained parameter space is clearly smaller than the one coming from the most restrictive constraint from the $\mu \to e \gamma$ branching ratio. Still, some points are removed from the $\alpha_{22}$ restriction, as shown in the top-middle panel, where the red points represent the parameter space excluded by the $\alpha_{22}$ limit. From this top-middle panel, we can also notice that the 
$Br(\mu-e \gamma)$ restriction pushes the $\alpha_{21}$ parameter up to values as low as few times $10^{-5}$. On the other hand, in the branching ratios $\tau \to e \gamma$ and $\tau \to \mu \gamma$, we see that the parameter space excluded by the non-unitary constraint is more significant than for the $\mu \to e \gamma$ case. To be more specific, in the $\tau \rightarrow e\gamma$,second row of Fig.~\ref{muon_to_electron}, we see that the restrictions on $\alpha_{31}$ rule out almost the same parameter space as the current branching ratio limits. Finally, for the $\tau \to \mu \gamma$ process, third row of Fig.~\ref{muon_to_electron}, we see a similar behavior as in the last case, with the difference that the $\alpha_{32}$ limit is more restrictive than the current limit of the process. As a consequence, we have another non-unitary parameter that could help to restrict the parameter space of the process. However, almost the same parameter space that is ruled out by $\alpha_{32}$ and allowed by the current limit is already ruled out by the $\alpha_{22}$.

We can have an additional view of this information by projecting the points under study on other planes, as the ones shown in Fig.~\ref{tau_to_electron}. 
This allows to get a better understanding of our analysis. For example, the top middle panel of this figure shows the blue points representing the excluded parameter space due to the $Br(\mu\to e\gamma)$ current limit, making evident that this limit is stronger than the $\alpha_{22}$ restriction. This figure also makes more evident that the $\mu\to e\gamma$ process points to a restriction in $\alpha_{21}$ of the order of $10^{-5}$ (top-left panel). Moreover, we also show in this figure the role of the non-unitary $\alpha_{22}$ limit in restricting the parameter space, as can be seen in the panels of the third column. Also, for this third column, we see that the excluded region by $\alpha_{22}$ tends disfavors values for the heavy sterile neutrinos below 1 TeV.

 \section{Conclusion}\label{Conclusions}
 
We studied the non-unitary effects on the inverse and linear seesaw models when the cLFV processes are highly suppressed. We also studied the complementarity between the non-unitarity effects and the cLFV processes in the inverse seesaw with one lepton flavor violating process at a time. We found that, as was expected, the $\mu\rightarrow e \gamma$ process gives the most stringent bound on the heavy sterile masses. In the other two cases, $\tau\rightarrow e \gamma$ and $\tau\rightarrow \mu \gamma$, the cLFV process and the $\alpha_{22}$ unitarity violating parameter are complementary. This complementarity disfavors masses of the heavy sterile states below 1 TeV. It is worth mentioning that for all the cases discussed, we provided a model based on the anomaly-free $B-L$ and a $Z_7$ ($Z_5$) flavor symmetry for the linear (inverse) seesaw where the light neutrino mass matrix presents one of the two-zero texture neutrino mass matrix in agreement with the current neutrino oscillation experiments.

\section*{Acknowledgements}

This work was partially supported by SNII-M\'exico, the CONAHCyT research Grant No. A1-S-23238 and the DGAPA UNAM grant PAPIIT IN111625. EP is grateful for the support of PASPA-DGAPA, UNAM for a sabbatical leave, and Fundaci\'on Marcos Moshinsky. JMCR acknowledges CONAHCyT for funding his PhD studies.

\appendix

\section{Block matrix diagonalization method (BMDM) for the type-I seesaw family}

In these appendices, we present the relevant formulas employed in our analysis, including the BMDM for the family of seesaw type-I models, the non-unitary effects (quantified by the $\eta$ matrix), and the formula for the branching ratio of the $\ell\to\ell'\gamma$ cLFV processes. 

For a model belonging to the type-I family seesaw, which besides the 3 light-active neutrinos, incorporates $m$ new heavy neutral fields, the neutrino mass matrix $M_\nu$ can be written as 
\begin{equation}
    M_{\nu_{n\times n}}=\begin{pmatrix}
0_{3\times 3} & M_{D_{3\times m}} \\ 
M^T_{D_{m \times 3}} & M_{R_{m \times m}}
\end{pmatrix},
\end{equation}
where $n=m+3$. We can get the physical masses by the following transformation 
\begin{equation}
\label{transformation}
    U^T M_{n\times n}U=M_{diag}.
\end{equation}
For the matrix $U$, we can consider the polar decomposition
\begin{equation}
\label{unitary matrix}
    U=exp(iH) \cdot V,
\end{equation}
where
\begin{equation}
 H=\begin{pmatrix}
0 & S \\ 
S^{\dagger} &0
\end{pmatrix}, \quad V= \begin{pmatrix}
V_{1} & 0 \\ 
0 & V_{2} 
\end{pmatrix}.       
\end{equation}
We can obtain a relation between S and the mass matrices from Eq. (\ref{transformation})
\begin{equation}
\label{S}
    iS^{*}=-M_D M^{-1}_{R}.
\end{equation}

Expanding Eq. (\ref{unitary matrix}) to second order on $S$ 
\begin{equation}
\label{U_masses}
U_{n \times n}= \begin{pmatrix}
\mathbb{I}_{3 \times 3}-\frac{1}{2}(M_D^*(M^*_R)^{-1}M^{-1}_RM^{T}_D)_{3\times 3} & (M^*_D(M_R^
*)^{-1})_{3\times m} \\ 
(M_R^{-1}M^T_D)_{m \times 3} &  \mathbb{I}_{3 \times 3}-\frac{1}{2}(M^{-1}_RM^{T}_DM_D^*(M^*_R)^{-1})_{m\times m}
\end{pmatrix} \cdot V.
\end{equation}
With this, the physical light and heavy neutrino masses are given by
\begin{align}
    m_{diag}=&-(V_1^T(M_DM^{-1}_RM^{T}_D)V_1)_{3\times 3}\\ 
    M^{diag}_{N}=& (V_2^TM_{R}V_2)_{m\times m},
\end{align}
where $V_1$ and $V_2$ diagonalize the light and heavy sectors, respectively. 

\subsection{Non-unitary effects}

Commonly, the non-unitary effects are searched in the oscillation experiments, searching an anomaly in the number of neutrino species that are detected ~\cite{Dutta:2019hmb,Denton_2022,Blennow:2023mqx,Gronau:1984ct,Nardi:1994iv,Atre:2009rg,Fernandez-Martinez:2016lgt,Celestino-Ramirez:2023zox,Escrihuela:2015wra,CentellesChulia:2024sff}. However, this is not the only way to study the non-unitary effects. In this appendix we will explain how to understand the non-unitary effects using the cLFV processes. To study the non-unitary effects using the cFLV processes, we need to understand the charged current lagrangian, which is described by: 
\begin{equation}
\mathcal{L}=-\frac{g}{\sqrt{2}}W^{-}_{\mu}\sum^{3}_{i=1}\sum^{n}_{j=1}K_{ij}\Bar{l}_i\gamma^{\mu}P_L \nu_j +h.c.,
\end{equation}
where $K_{ij}$ generally is a rectangular matrix:
\begin{equation}
K_{ij}=\sum^{3}_{k=1}U^{\ell *}_{ki}U^{\nu}_{kj} .
\end{equation}
The dimension of the matrix depends on the number of new massive states. In general, we can say that the K matrix has $3\times n$ dimension. We will work in the diagonal basis of the charged lepton mixing matrix ($U^{\ell}_{ik}=\delta_{ik}$). It is helpful to describe the neutrino mixing matrix as a block matrix: 
\begin{equation}
K=(K_{L_{3\times 3}},K_{H_{3\times m}}),
\end{equation}
where $K_L$ and $K_H$ are the light and heavy sectors respectively. We can extract these matrices from Eq. (\ref{U_masses})
\begin{align}
\label{non-unitarypart} K_{L_{3\times 3}}=&(\mathbb{I}_{3 \times 3}-\frac{1}{2}(M_D^*(M^*_R)^{-1}M^{-1}_RM^{T}_D)_{3\times 3})V_{1_{3\times 3}}\\
K_{H_{3\times m}}=& (M^*_D(M_R^*)^{-1})_{3\times m}V_{2_{m \times m}}.
\end{align}
We see in Eq. (\ref{non-unitarypart}) a deviation from the light unitary mixing matrix $V_1$:
\begin{equation}
K_{L_{3\times 3}}=(\mathbb{I}_{3 \times 3}-\eta)V_{1_{3\times 3}},
\end{equation}
where
\begin{equation}
\label{non_unitary_effect-O}
\eta=\frac{1}{2}(M_D^*(M^*_R)^{-1}M^{-1}_RM^{T}_D)_{3\times 3}.\end{equation}

Now, by considering the neutrino mass matrix given in Eq. (\ref{inverse seesaw}) [(\ref{Linear seesaw})] for the inverse [linear] model, and by using the BMDM, in the  $\mu\to 0$ [$M_L\to 0$] limit, it is straightforward to derive the expressions for the $m_\nu$, $\eta$, $K_L$, and $K_H$ matrices presented in Table \ref{R-formulas}.

\begin{table}[h!]
\begin{tabular}{ccc}
Matrix & \multicolumn{1}{c}{Inverse Seesaw} & Linear Seesaw \\ \hline \hline
$M_\nu$  & \multicolumn{1}{|c|}{ 
$
\begin{pmatrix}
0 & M_D & 0\\ 
M^T_D & 0 & M\\
0 & M^T  & \mu
\end{pmatrix},\quad$ \tagarray\label{inverse seesaw}
}  &      
$
\begin{pmatrix}
0 & M_D & M_L\\ 
M^T_D & 0 & M\\
M^T_L & M^T  & 0
\end{pmatrix},\quad
$ \tagarray\label{Linear seesaw}
\\ \hline
$m_\nu$ & \multicolumn{1}{|c|}{
$
(M_D(M^T)^{-1}\mu M^{-1}M^{T}_D)_{3\times 3},\quad$ \tagarray\label{mass_inverseseesaw}
}               &   
$M_D(M_LM^{-1})^T +(M_LM^{-1})M^T_D,\quad$ \tagarray\label{linear_mass}
\\ \hline
$\eta_{3\times 3}$                 & \multicolumn{2}{c}{$
\frac{1}{2}(M_D^*(M^*)^{-1}M^{-1}M^{T}_D)_{3\times 3},\quad     
$\tagarray\label{non_unitary_effect}}                             \\ \hline
$K_{L_{3\times 3}}$ & \multicolumn{2}{c}{
$(\mathbb{I}_{3 \times 3}-\eta)V_{1_{3\times 3}},\quad$ \tagarray\label{KL}
}                               \\ \hline
$K_{H_{3\times 6}}$ & \multicolumn{2}{c}{ $
\bigg(0_{3\times 3}, (M_D^* (M^{*T})^{-1})_{3\times 3}\bigg)\cdot V_{2_{6\times 6}},\quad
$ \tagarray\label{KH} }                           \\ \hline \hline
\end{tabular}
\caption{\label{BMDM-ls-formulas}Relevant matrices for the low-scale seesaw models using the BMDM.}
\label{R-formulas}
\end{table}


\section{\pdfmath{\ell \to \ell^{\prime} \gamma} cLFV decay}

The branching ratio of these processes, neglecting the mass of the lighter-charged lepton, are given by \cite{He:2002pva,Ilakovac:1994kj,Hernandez-Tome:2019lkb}:
\begin{align}
    BR(\ell \to \ell^{\prime} \gamma) &=\frac{\alpha}{\gamma_{\ell}}m^3_{\ell}|F^{\gamma}_M (0)|^2, \\
    F^{\gamma}_M (0) &=\frac{\alpha_W}{16 \pi}\frac{m_{\ell}}{M^2_W}\sum_i K^{*}_{\ell i}K_{\ell^{\prime} i} f^{\gamma}_M(x_i),\notag \\ 
    f^{\gamma}_M(x_i) &=\frac{3x^3 logx}{2(x-1)^4}-\frac{2x^3 +5x^2 -x}{4(x-1)^3}+\frac{5}{6}, \notag
\end{align}
where $\alpha= e^2/4\pi$ is the fine-structure constant, and we have defined $\alpha_W\equiv \alpha /s^2_W$, $x_i \equiv m^2_{\chi_i}/M^2_W$, with $m_{\chi_i}$ is the physical masses of the neutrino states. The current and future expected limits for these branching ratios are listed in Table~\ref{BRlimits}.

\begin{table}[h!]
\def\arraystretch{2}
\begin{tabular}{|c|c|c|}
\hline
Process & Present limit & Future Sensitivity \\ \hline
$\mu \to e \gamma$       & $4.2 \times 10^{-13}$ \cite{MEG:2013oxv}         & $6 \times 10^{-14}$ \cite{MEGII:2018kmf}                  \\ \hline
$\tau \to e \gamma$      & $3.3 \times 10^{-8}$ \cite{BaBar:2009hkt}          & $3 \times 10^{-9}$\cite{Belle-II:2018jsg}                  \\ \hline
$\tau \to \mu \gamma$       & $4.2 \times 10^{-8}$ \cite{BaBar:2009hkt}         & $10^{-9}$  \cite{Belle-II:2018jsg}                \\ \hline
\end{tabular}
\caption{\label{table cLFV limits} Present and future sensitivities for the $\ell\to\ell'\gamma$ cLFV decays.}.
\label{BRlimits}
\end{table}

\bibliography{Reference}

\begin{thebibliography}{42}%
\makeatletter
\providecommand \@ifxundefined [1]{%
 \@ifx{#1\undefined}
}%
\providecommand \@ifnum [1]{%
 \ifnum #1\expandafter \@firstoftwo
 \else \expandafter \@secondoftwo
 \fi
}%
\providecommand \@ifx [1]{%
 \ifx #1\expandafter \@firstoftwo
 \else \expandafter \@secondoftwo
 \fi
}%
\providecommand \natexlab [1]{#1}%
\providecommand \enquote  [1]{``#1''}%
\providecommand \bibnamefont  [1]{#1}%
\providecommand \bibfnamefont [1]{#1}%
\providecommand \citenamefont [1]{#1}%
\providecommand \href@noop [0]{\@secondoftwo}%
\providecommand \href [0]{\begingroup \@sanitize@url \@href}%
\providecommand \@href[1]{\@@startlink{#1}\@@href}%
\providecommand \@@href[1]{\endgroup#1\@@endlink}%
\providecommand \@sanitize@url [0]{\catcode `\\12\catcode `\$12\catcode `\&12\catcode `\#12\catcode `\^12\catcode `\_12\catcode `\%12\relax}%
\providecommand \@@startlink[1]{}%
\providecommand \@@endlink[0]{}%
\providecommand \url  [0]{\begingroup\@sanitize@url \@url }%
\providecommand \@url [1]{\endgroup\@href {#1}{\urlprefix }}%
\providecommand \urlprefix  [0]{URL }%
\providecommand \Eprint [0]{\href }%
\providecommand \doibase [0]{http://dx.doi.org/}%
\providecommand \selectlanguage [0]{\@gobble}%
\providecommand \bibinfo  [0]{\@secondoftwo}%
\providecommand \bibfield  [0]{\@secondoftwo}%
\providecommand \translation [1]{[#1]}%
\providecommand \BibitemOpen [0]{}%
\providecommand \bibitemStop [0]{}%
\providecommand \bibitemNoStop [0]{.\EOS\space}%
\providecommand \EOS [0]{\spacefactor3000\relax}%
\providecommand \BibitemShut  [1]{\csname bibitem#1\endcsname}%
\let\auto@bib@innerbib\@empty
\bibitem [{\citenamefont {Schechter}\ and\ \citenamefont {Valle}(1982)}]{Schechter:1981cv}%
  \BibitemOpen
  \bibfield  {author} {\bibinfo {author} {\bibfnamefont {J.}~\bibnamefont {Schechter}}\ and\ \bibinfo {author} {\bibfnamefont {J.~W.~F.}\ \bibnamefont {Valle}},\ }\href {\doibase 10.1103/PhysRevD.25.774} {\bibfield  {journal} {\bibinfo  {journal} {Phys. Rev. D}\ }\textbf {\bibinfo {volume} {25}},\ \bibinfo {pages} {774} (\bibinfo {year} {1982})}\BibitemShut {NoStop}%
\bibitem [{\citenamefont {Mohapatra}\ and\ \citenamefont {Valle}(1986)}]{Mohapatra:1986bd}%
  \BibitemOpen
  \bibfield  {author} {\bibinfo {author} {\bibfnamefont {R.~N.}\ \bibnamefont {Mohapatra}}\ and\ \bibinfo {author} {\bibfnamefont {J.~W.~F.}\ \bibnamefont {Valle}},\ }\href {\doibase 10.1103/PhysRevD.34.1642} {\bibfield  {journal} {\bibinfo  {journal} {Phys. Rev. D}\ }\textbf {\bibinfo {volume} {34}},\ \bibinfo {pages} {1642} (\bibinfo {year} {1986})}\BibitemShut {NoStop}%
\bibitem [{\citenamefont {Minkowski}(1977)}]{Minkowski:1977sc}%
  \BibitemOpen
  \bibfield  {author} {\bibinfo {author} {\bibfnamefont {P.}~\bibnamefont {Minkowski}},\ }\href {\doibase 10.1016/0370-2693(77)90435-X} {\bibfield  {journal} {\bibinfo  {journal} {Phys. Lett. B}\ }\textbf {\bibinfo {volume} {67}},\ \bibinfo {pages} {421} (\bibinfo {year} {1977})}\BibitemShut {NoStop}%
\bibitem [{\citenamefont {Yanagida}(1979)}]{Yanagida:1979as}%
  \BibitemOpen
  \bibfield  {author} {\bibinfo {author} {\bibfnamefont {T.}~\bibnamefont {Yanagida}},\ }\href@noop {} {\bibfield  {journal} {\bibinfo  {journal} {Conf. Proc. C}\ }\textbf {\bibinfo {volume} {7902131}},\ \bibinfo {pages} {95} (\bibinfo {year} {1979})}\BibitemShut {NoStop}%
\bibitem [{\citenamefont {Foot}\ \emph {et~al.}(1989)\citenamefont {Foot}, \citenamefont {Lew}, \citenamefont {He},\ and\ \citenamefont {Joshi}}]{Foot:1988aq}%
  \BibitemOpen
  \bibfield  {author} {\bibinfo {author} {\bibfnamefont {R.}~\bibnamefont {Foot}}, \bibinfo {author} {\bibfnamefont {H.}~\bibnamefont {Lew}}, \bibinfo {author} {\bibfnamefont {X.~G.}\ \bibnamefont {He}}, \ and\ \bibinfo {author} {\bibfnamefont {G.~C.}\ \bibnamefont {Joshi}},\ }\href {\doibase 10.1007/BF01415558} {\bibfield  {journal} {\bibinfo  {journal} {Z. Phys. C}\ }\textbf {\bibinfo {volume} {44}},\ \bibinfo {pages} {441} (\bibinfo {year} {1989})}\BibitemShut {NoStop}%
\bibitem [{\citenamefont {Cheng}\ and\ \citenamefont {Li}(1980)}]{Cheng:1980qt}%
  \BibitemOpen
  \bibfield  {author} {\bibinfo {author} {\bibfnamefont {T.~P.}\ \bibnamefont {Cheng}}\ and\ \bibinfo {author} {\bibfnamefont {L.-F.}\ \bibnamefont {Li}},\ }\href {\doibase 10.1103/PhysRevD.22.2860} {\bibfield  {journal} {\bibinfo  {journal} {Phys. Rev. D}\ }\textbf {\bibinfo {volume} {22}},\ \bibinfo {pages} {2860} (\bibinfo {year} {1980})}\BibitemShut {NoStop}%
\bibitem [{\citenamefont {Malinsky}\ \emph {et~al.}(2005)\citenamefont {Malinsky}, \citenamefont {Romao},\ and\ \citenamefont {Valle}}]{Malinsky:2005bi}%
  \BibitemOpen
  \bibfield  {author} {\bibinfo {author} {\bibfnamefont {M.}~\bibnamefont {Malinsky}}, \bibinfo {author} {\bibfnamefont {J.~C.}\ \bibnamefont {Romao}}, \ and\ \bibinfo {author} {\bibfnamefont {J.~W.~F.}\ \bibnamefont {Valle}},\ }\href {\doibase 10.1103/PhysRevLett.95.161801} {\bibfield  {journal} {\bibinfo  {journal} {Phys. Rev. Lett.}\ }\textbf {\bibinfo {volume} {95}},\ \bibinfo {pages} {161801} (\bibinfo {year} {2005})},\ \Eprint {http://arxiv.org/abs/hep-ph/0506296} {arXiv:hep-ph/0506296} \BibitemShut {NoStop}%
\bibitem [{\citenamefont {Gonzalez-Garcia}\ and\ \citenamefont {Valle}(1989)}]{Gonzalez-Garcia:1988okv}%
  \BibitemOpen
  \bibfield  {author} {\bibinfo {author} {\bibfnamefont {M.~C.}\ \bibnamefont {Gonzalez-Garcia}}\ and\ \bibinfo {author} {\bibfnamefont {J.~W.~F.}\ \bibnamefont {Valle}},\ }\href {\doibase 10.1016/0370-2693(89)91131-3} {\bibfield  {journal} {\bibinfo  {journal} {Phys. Lett. B}\ }\textbf {\bibinfo {volume} {216}},\ \bibinfo {pages} {360} (\bibinfo {year} {1989})}\BibitemShut {NoStop}%
\bibitem [{\citenamefont {Centelles~Chuli\'a}\ \emph {et~al.}(2024{\natexlab{a}})\citenamefont {Centelles~Chuli\'a}, \citenamefont {Herrero-Brocal},\ and\ \citenamefont {Vicente}}]{CentellesChulia:2024uzv}%
  \BibitemOpen
  \bibfield  {author} {\bibinfo {author} {\bibfnamefont {S.}~\bibnamefont {Centelles~Chuli\'a}}, \bibinfo {author} {\bibfnamefont {A.}~\bibnamefont {Herrero-Brocal}}, \ and\ \bibinfo {author} {\bibfnamefont {A.}~\bibnamefont {Vicente}},\ }\href@noop {} {\  (\bibinfo {year} {2024}{\natexlab{a}})},\ \Eprint {http://arxiv.org/abs/2404.15415} {arXiv:2404.15415 [hep-ph]} \BibitemShut {NoStop}%
\bibitem [{\citenamefont {Garnica}\ \emph {et~al.}(2023)\citenamefont {Garnica}, \citenamefont {Hern\'andez-Tom\'e},\ and\ \citenamefont {Peinado}}]{Garnica:2023ccx}%
  \BibitemOpen
  \bibfield  {author} {\bibinfo {author} {\bibfnamefont {J.~C.}\ \bibnamefont {Garnica}}, \bibinfo {author} {\bibfnamefont {G.}~\bibnamefont {Hern\'andez-Tom\'e}}, \ and\ \bibinfo {author} {\bibfnamefont {E.}~\bibnamefont {Peinado}},\ }\href {\doibase 10.1103/PhysRevD.108.035033} {\bibfield  {journal} {\bibinfo  {journal} {Phys. Rev. D}\ }\textbf {\bibinfo {volume} {108}},\ \bibinfo {pages} {035033} (\bibinfo {year} {2023})},\ \Eprint {http://arxiv.org/abs/2302.07379} {arXiv:2302.07379 [hep-ph]} \BibitemShut {NoStop}%
\bibitem [{\citenamefont {Arganda}\ \emph {et~al.}(2015)\citenamefont {Arganda}, \citenamefont {Herrero}, \citenamefont {Marcano},\ and\ \citenamefont {Weiland}}]{Arganda:2014dta}%
  \BibitemOpen
  \bibfield  {author} {\bibinfo {author} {\bibfnamefont {E.}~\bibnamefont {Arganda}}, \bibinfo {author} {\bibfnamefont {M.~J.}\ \bibnamefont {Herrero}}, \bibinfo {author} {\bibfnamefont {X.}~\bibnamefont {Marcano}}, \ and\ \bibinfo {author} {\bibfnamefont {C.}~\bibnamefont {Weiland}},\ }\href {\doibase 10.1103/PhysRevD.91.015001} {\bibfield  {journal} {\bibinfo  {journal} {Phys. Rev. D}\ }\textbf {\bibinfo {volume} {91}},\ \bibinfo {pages} {015001} (\bibinfo {year} {2015})},\ \Eprint {http://arxiv.org/abs/1405.4300} {arXiv:1405.4300 [hep-ph]} \BibitemShut {NoStop}%
\bibitem [{\citenamefont {de~Salas}\ \emph {et~al.}(2021)\citenamefont {de~Salas}, \citenamefont {Forero}, \citenamefont {Gariazzo}, \citenamefont {Mart\'\i{}nez-Mirav\'e}, \citenamefont {Mena}, \citenamefont {Ternes}, \citenamefont {T\'ortola},\ and\ \citenamefont {Valle}}]{deSalas:2020pgw}%
  \BibitemOpen
  \bibfield  {author} {\bibinfo {author} {\bibfnamefont {P.~F.}\ \bibnamefont {de~Salas}}, \bibinfo {author} {\bibfnamefont {D.~V.}\ \bibnamefont {Forero}}, \bibinfo {author} {\bibfnamefont {S.}~\bibnamefont {Gariazzo}}, \bibinfo {author} {\bibfnamefont {P.}~\bibnamefont {Mart\'\i{}nez-Mirav\'e}}, \bibinfo {author} {\bibfnamefont {O.}~\bibnamefont {Mena}}, \bibinfo {author} {\bibfnamefont {C.~A.}\ \bibnamefont {Ternes}}, \bibinfo {author} {\bibfnamefont {M.}~\bibnamefont {T\'ortola}}, \ and\ \bibinfo {author} {\bibfnamefont {J.~W.~F.}\ \bibnamefont {Valle}},\ }\href {\doibase 10.1007/JHEP02(2021)071} {\bibfield  {journal} {\bibinfo  {journal} {JHEP}\ }\textbf {\bibinfo {volume} {02}},\ \bibinfo {pages} {071} (\bibinfo {year} {2021})},\ \Eprint {http://arxiv.org/abs/2006.11237} {arXiv:2006.11237 [hep-ph]} \BibitemShut {NoStop}%
\bibitem [{\citenamefont {Escrihuela}\ \emph {et~al.}(2015)\citenamefont {Escrihuela}, \citenamefont {Forero}, \citenamefont {Miranda}, \citenamefont {Tortola},\ and\ \citenamefont {Valle}}]{Escrihuela:2015wra}%
  \BibitemOpen
  \bibfield  {author} {\bibinfo {author} {\bibfnamefont {F.~J.}\ \bibnamefont {Escrihuela}}, \bibinfo {author} {\bibfnamefont {D.~V.}\ \bibnamefont {Forero}}, \bibinfo {author} {\bibfnamefont {O.~G.}\ \bibnamefont {Miranda}}, \bibinfo {author} {\bibfnamefont {M.}~\bibnamefont {Tortola}}, \ and\ \bibinfo {author} {\bibfnamefont {J.~W.~F.}\ \bibnamefont {Valle}},\ }\href {\doibase 10.1103/PhysRevD.92.053009} {\bibfield  {journal} {\bibinfo  {journal} {Phys. Rev. D}\ }\textbf {\bibinfo {volume} {92}},\ \bibinfo {pages} {053009} (\bibinfo {year} {2015})},\ \bibinfo {note} {[Erratum: Phys.Rev.D 93, 119905 (2016)]},\ \Eprint {http://arxiv.org/abs/1503.08879} {arXiv:1503.08879 [hep-ph]} \BibitemShut {NoStop}%
\bibitem [{\citenamefont {Celestino-Ram\'\i{}rez}\ \emph {et~al.}(2024)\citenamefont {Celestino-Ram\'\i{}rez}, \citenamefont {Escrihuela}, \citenamefont {Flores},\ and\ \citenamefont {Miranda}}]{Celestino-Ramirez:2023zox}%
  \BibitemOpen
  \bibfield  {author} {\bibinfo {author} {\bibfnamefont {J.~M.}\ \bibnamefont {Celestino-Ram\'\i{}rez}}, \bibinfo {author} {\bibfnamefont {F.~J.}\ \bibnamefont {Escrihuela}}, \bibinfo {author} {\bibfnamefont {L.~J.}\ \bibnamefont {Flores}}, \ and\ \bibinfo {author} {\bibfnamefont {O.~G.}\ \bibnamefont {Miranda}},\ }\href {\doibase 10.1103/PhysRevD.109.L011705} {\bibfield  {journal} {\bibinfo  {journal} {Phys. Rev. D}\ }\textbf {\bibinfo {volume} {109}},\ \bibinfo {pages} {L011705} (\bibinfo {year} {2024})},\ \Eprint {http://arxiv.org/abs/2309.00116} {arXiv:2309.00116 [hep-ph]} \BibitemShut {NoStop}%
\bibitem [{\citenamefont {Chatterjee}\ \emph {et~al.}(2022)\citenamefont {Chatterjee}, \citenamefont {Miranda}, \citenamefont {T\'ortola},\ and\ \citenamefont {Valle}}]{Chatterjee:2021xyu}%
  \BibitemOpen
  \bibfield  {author} {\bibinfo {author} {\bibfnamefont {S.~S.}\ \bibnamefont {Chatterjee}}, \bibinfo {author} {\bibfnamefont {O.~G.}\ \bibnamefont {Miranda}}, \bibinfo {author} {\bibfnamefont {M.}~\bibnamefont {T\'ortola}}, \ and\ \bibinfo {author} {\bibfnamefont {J.~W.~F.}\ \bibnamefont {Valle}},\ }\href {\doibase 10.1103/PhysRevD.106.075016} {\bibfield  {journal} {\bibinfo  {journal} {Phys. Rev. D}\ }\textbf {\bibinfo {volume} {106}},\ \bibinfo {pages} {075016} (\bibinfo {year} {2022})},\ \Eprint {http://arxiv.org/abs/2111.08673} {arXiv:2111.08673 [hep-ph]} \BibitemShut {NoStop}%
\bibitem [{\citenamefont {Centelles~Chuli\'a}\ \emph {et~al.}(2024{\natexlab{b}})\citenamefont {Centelles~Chuli\'a}, \citenamefont {Miranda},\ and\ \citenamefont {Valle}}]{CentellesChulia:2024sff}%
  \BibitemOpen
  \bibfield  {author} {\bibinfo {author} {\bibfnamefont {S.}~\bibnamefont {Centelles~Chuli\'a}}, \bibinfo {author} {\bibfnamefont {O.}~\bibnamefont {Miranda}}, \ and\ \bibinfo {author} {\bibfnamefont {J.~W.~F.}\ \bibnamefont {Valle}},\ }\href@noop {} {\  (\bibinfo {year} {2024}{\natexlab{b}})},\ \Eprint {http://arxiv.org/abs/2402.00114} {arXiv:2402.00114 [hep-ph]} \BibitemShut {NoStop}%
\bibitem [{\citenamefont {Forero}\ \emph {et~al.}(2011)\citenamefont {Forero}, \citenamefont {Morisi}, \citenamefont {Tortola},\ and\ \citenamefont {Valle}}]{Forero:2011pc}%
  \BibitemOpen
  \bibfield  {author} {\bibinfo {author} {\bibfnamefont {D.~V.}\ \bibnamefont {Forero}}, \bibinfo {author} {\bibfnamefont {S.}~\bibnamefont {Morisi}}, \bibinfo {author} {\bibfnamefont {M.}~\bibnamefont {Tortola}}, \ and\ \bibinfo {author} {\bibfnamefont {J.~W.~F.}\ \bibnamefont {Valle}},\ }\href {\doibase 10.1007/JHEP09(2011)142} {\bibfield  {journal} {\bibinfo  {journal} {JHEP}\ }\textbf {\bibinfo {volume} {09}},\ \bibinfo {pages} {142} (\bibinfo {year} {2011})},\ \Eprint {http://arxiv.org/abs/1107.6009} {arXiv:1107.6009 [hep-ph]} \BibitemShut {NoStop}%
\bibitem [{\citenamefont {Batra}\ \emph {et~al.}(2023)\citenamefont {Batra}, \citenamefont {Bharadwaj}, \citenamefont {Mandal}, \citenamefont {Srivastava},\ and\ \citenamefont {Valle}}]{Batra:2023mds}%
  \BibitemOpen
  \bibfield  {author} {\bibinfo {author} {\bibfnamefont {A.}~\bibnamefont {Batra}}, \bibinfo {author} {\bibfnamefont {P.}~\bibnamefont {Bharadwaj}}, \bibinfo {author} {\bibfnamefont {S.}~\bibnamefont {Mandal}}, \bibinfo {author} {\bibfnamefont {R.}~\bibnamefont {Srivastava}}, \ and\ \bibinfo {author} {\bibfnamefont {J.~W.~F.}\ \bibnamefont {Valle}},\ }\href {\doibase 10.1007/JHEP07(2023)221} {\bibfield  {journal} {\bibinfo  {journal} {JHEP}\ }\textbf {\bibinfo {volume} {07}},\ \bibinfo {pages} {221} (\bibinfo {year} {2023})},\ \Eprint {http://arxiv.org/abs/2305.00994} {arXiv:2305.00994 [hep-ph]} \BibitemShut {NoStop}%
\bibitem [{\citenamefont {Frampton}\ \emph {et~al.}(2002)\citenamefont {Frampton}, \citenamefont {Glashow},\ and\ \citenamefont {Marfatia}}]{Frampton:2002yf}%
  \BibitemOpen
  \bibfield  {author} {\bibinfo {author} {\bibfnamefont {P.~H.}\ \bibnamefont {Frampton}}, \bibinfo {author} {\bibfnamefont {S.~L.}\ \bibnamefont {Glashow}}, \ and\ \bibinfo {author} {\bibfnamefont {D.}~\bibnamefont {Marfatia}},\ }\href {\doibase 10.1016/S0370-2693(02)01817-8} {\bibfield  {journal} {\bibinfo  {journal} {Phys. Lett. B}\ }\textbf {\bibinfo {volume} {536}},\ \bibinfo {pages} {79} (\bibinfo {year} {2002})},\ \Eprint {http://arxiv.org/abs/hep-ph/0201008} {arXiv:hep-ph/0201008} \BibitemShut {NoStop}%
\bibitem [{\citenamefont {Ludl}\ \emph {et~al.}(2012)\citenamefont {Ludl}, \citenamefont {Morisi},\ and\ \citenamefont {Peinado}}]{Ludl:2011vv}%
  \BibitemOpen
  \bibfield  {author} {\bibinfo {author} {\bibfnamefont {P.~O.}\ \bibnamefont {Ludl}}, \bibinfo {author} {\bibfnamefont {S.}~\bibnamefont {Morisi}}, \ and\ \bibinfo {author} {\bibfnamefont {E.}~\bibnamefont {Peinado}},\ }\href {\doibase 10.1016/j.nuclphysb.2011.12.017} {\bibfield  {journal} {\bibinfo  {journal} {Nucl. Phys. B}\ }\textbf {\bibinfo {volume} {857}},\ \bibinfo {pages} {411} (\bibinfo {year} {2012})},\ \Eprint {http://arxiv.org/abs/1109.3393} {arXiv:1109.3393 [hep-ph]} \BibitemShut {NoStop}%
\bibitem [{\citenamefont {Meloni}\ \emph {et~al.}(2014)\citenamefont {Meloni}, \citenamefont {Meroni},\ and\ \citenamefont {Peinado}}]{Meloni:2014yea}%
  \BibitemOpen
  \bibfield  {author} {\bibinfo {author} {\bibfnamefont {D.}~\bibnamefont {Meloni}}, \bibinfo {author} {\bibfnamefont {A.}~\bibnamefont {Meroni}}, \ and\ \bibinfo {author} {\bibfnamefont {E.}~\bibnamefont {Peinado}},\ }\href {\doibase 10.1103/PhysRevD.89.053009} {\bibfield  {journal} {\bibinfo  {journal} {Phys. Rev. D}\ }\textbf {\bibinfo {volume} {89}},\ \bibinfo {pages} {053009} (\bibinfo {year} {2014})},\ \Eprint {http://arxiv.org/abs/1401.3207} {arXiv:1401.3207 [hep-ph]} \BibitemShut {NoStop}%
\bibitem [{\citenamefont {De~La~Vega}\ \emph {et~al.}(2019)\citenamefont {De~La~Vega}, \citenamefont {Ferro-Hernandez},\ and\ \citenamefont {Peinado}}]{DeLaVega:2018bkp}%
  \BibitemOpen
  \bibfield  {author} {\bibinfo {author} {\bibfnamefont {L.~M.~G.}\ \bibnamefont {De~La~Vega}}, \bibinfo {author} {\bibfnamefont {R.}~\bibnamefont {Ferro-Hernandez}}, \ and\ \bibinfo {author} {\bibfnamefont {E.}~\bibnamefont {Peinado}},\ }\href {\doibase 10.1103/PhysRevD.99.055044} {\bibfield  {journal} {\bibinfo  {journal} {Phys. Rev. D}\ }\textbf {\bibinfo {volume} {99}},\ \bibinfo {pages} {055044} (\bibinfo {year} {2019})},\ \Eprint {http://arxiv.org/abs/1811.10619} {arXiv:1811.10619 [hep-ph]} \BibitemShut {NoStop}%
\bibitem [{\citenamefont {Alcaide}\ \emph {et~al.}(2018)\citenamefont {Alcaide}, \citenamefont {Salvado},\ and\ \citenamefont {Santamaria}}]{Alcaide:2018vni}%
  \BibitemOpen
  \bibfield  {author} {\bibinfo {author} {\bibfnamefont {J.}~\bibnamefont {Alcaide}}, \bibinfo {author} {\bibfnamefont {J.}~\bibnamefont {Salvado}}, \ and\ \bibinfo {author} {\bibfnamefont {A.}~\bibnamefont {Santamaria}},\ }\href {\doibase 10.1007/JHEP07(2018)164} {\bibfield  {journal} {\bibinfo  {journal} {JHEP}\ }\textbf {\bibinfo {volume} {07}},\ \bibinfo {pages} {164} (\bibinfo {year} {2018})},\ \Eprint {http://arxiv.org/abs/1806.06785} {arXiv:1806.06785 [hep-ph]} \BibitemShut {NoStop}%
\bibitem [{\citenamefont {Weinberg}(1979)}]{Weinberg:1979sa}%
  \BibitemOpen
  \bibfield  {author} {\bibinfo {author} {\bibfnamefont {S.}~\bibnamefont {Weinberg}},\ }\href {\doibase 10.1103/PhysRevLett.43.1566} {\bibfield  {journal} {\bibinfo  {journal} {Phys. Rev. Lett.}\ }\textbf {\bibinfo {volume} {43}},\ \bibinfo {pages} {1566} (\bibinfo {year} {1979})}\BibitemShut {NoStop}%
\bibitem [{\citenamefont {Forero}\ \emph {et~al.}(2021)\citenamefont {Forero}, \citenamefont {Giunti}, \citenamefont {Ternes},\ and\ \citenamefont {Tortola}}]{Forero:2021azc}%
  \BibitemOpen
  \bibfield  {author} {\bibinfo {author} {\bibfnamefont {D.~V.}\ \bibnamefont {Forero}}, \bibinfo {author} {\bibfnamefont {C.}~\bibnamefont {Giunti}}, \bibinfo {author} {\bibfnamefont {C.~A.}\ \bibnamefont {Ternes}}, \ and\ \bibinfo {author} {\bibfnamefont {M.}~\bibnamefont {Tortola}},\ }\href {\doibase 10.1103/PhysRevD.104.075030} {\bibfield  {journal} {\bibinfo  {journal} {Phys. Rev. D}\ }\textbf {\bibinfo {volume} {104}},\ \bibinfo {pages} {075030} (\bibinfo {year} {2021})},\ \Eprint {http://arxiv.org/abs/2103.01998} {arXiv:2103.01998 [hep-ph]} \BibitemShut {NoStop}%
\bibitem [{\citenamefont {Rodejohann}\ and\ \citenamefont {Valle}(2011)}]{Rodejohann:2011vc}%
  \BibitemOpen
  \bibfield  {author} {\bibinfo {author} {\bibfnamefont {W.}~\bibnamefont {Rodejohann}}\ and\ \bibinfo {author} {\bibfnamefont {J.~W.~F.}\ \bibnamefont {Valle}},\ }\href {\doibase 10.1103/PhysRevD.84.073011} {\bibfield  {journal} {\bibinfo  {journal} {Phys. Rev. D}\ }\textbf {\bibinfo {volume} {84}},\ \bibinfo {pages} {073011} (\bibinfo {year} {2011})},\ \Eprint {http://arxiv.org/abs/1108.3484} {arXiv:1108.3484 [hep-ph]} \BibitemShut {NoStop}%
\bibitem [{\citenamefont {Blennow}\ \emph {et~al.}(2017)\citenamefont {Blennow}, \citenamefont {Coloma}, \citenamefont {Fernandez-Martinez}, \citenamefont {Hernandez-Garcia},\ and\ \citenamefont {Lopez-Pavon}}]{Blennow:2016jkn}%
  \BibitemOpen
  \bibfield  {author} {\bibinfo {author} {\bibfnamefont {M.}~\bibnamefont {Blennow}}, \bibinfo {author} {\bibfnamefont {P.}~\bibnamefont {Coloma}}, \bibinfo {author} {\bibfnamefont {E.}~\bibnamefont {Fernandez-Martinez}}, \bibinfo {author} {\bibfnamefont {J.}~\bibnamefont {Hernandez-Garcia}}, \ and\ \bibinfo {author} {\bibfnamefont {J.}~\bibnamefont {Lopez-Pavon}},\ }\href {\doibase 10.1007/JHEP04(2017)153} {\bibfield  {journal} {\bibinfo  {journal} {JHEP}\ }\textbf {\bibinfo {volume} {04}},\ \bibinfo {pages} {153} (\bibinfo {year} {2017})},\ \Eprint {http://arxiv.org/abs/1609.08637} {arXiv:1609.08637 [hep-ph]} \BibitemShut {NoStop}%
\bibitem [{\citenamefont {Celestino-Ramirez}\ and\ \citenamefont {Miranda}(2024)}]{Celestino-Ramirez:2024gmq}%
  \BibitemOpen
  \bibfield  {author} {\bibinfo {author} {\bibfnamefont {J.~M.}\ \bibnamefont {Celestino-Ramirez}}\ and\ \bibinfo {author} {\bibfnamefont {O.~G.}\ \bibnamefont {Miranda}},\ }\href@noop {} {\  (\bibinfo {year} {2024})},\ \Eprint {http://arxiv.org/abs/2405.03907} {arXiv:2405.03907 [hep-ph]} \BibitemShut {NoStop}%
\bibitem [{\citenamefont {Dutta}\ and\ \citenamefont {Roy}(2021)}]{Dutta:2019hmb}%
  \BibitemOpen
  \bibfield  {author} {\bibinfo {author} {\bibfnamefont {D.}~\bibnamefont {Dutta}}\ and\ \bibinfo {author} {\bibfnamefont {S.}~\bibnamefont {Roy}},\ }\href {\doibase 10.1088/1361-6471/abdc03} {\bibfield  {journal} {\bibinfo  {journal} {J. Phys. G}\ }\textbf {\bibinfo {volume} {48}},\ \bibinfo {pages} {045004} (\bibinfo {year} {2021})},\ \Eprint {http://arxiv.org/abs/1901.11298} {arXiv:1901.11298 [hep-ph]} \BibitemShut {NoStop}%
\bibitem [{\citenamefont {Denton}\ and\ \citenamefont {Gehrlein}(2022)}]{Denton_2022}%
  \BibitemOpen
  \bibfield  {author} {\bibinfo {author} {\bibfnamefont {P.~B.}\ \bibnamefont {Denton}}\ and\ \bibinfo {author} {\bibfnamefont {J.}~\bibnamefont {Gehrlein}},\ }\href {\doibase 10.1007/jhep06(2022)135} {\bibfield  {journal} {\bibinfo  {journal} {Journal of High Energy Physics}\ }\textbf {\bibinfo {volume} {2022}} (\bibinfo {year} {2022}),\ 10.1007/jhep06(2022)135}\BibitemShut {NoStop}%
\bibitem [{\citenamefont {Blennow}\ \emph {et~al.}(2023)\citenamefont {Blennow}, \citenamefont {Fern\'andez-Mart\'\i{}nez}, \citenamefont {Hern\'andez-Garc\'\i{}a}, \citenamefont {L\'opez-Pav\'on}, \citenamefont {Marcano},\ and\ \citenamefont {Naredo-Tuero}}]{Blennow:2023mqx}%
  \BibitemOpen
  \bibfield  {author} {\bibinfo {author} {\bibfnamefont {M.}~\bibnamefont {Blennow}}, \bibinfo {author} {\bibfnamefont {E.}~\bibnamefont {Fern\'andez-Mart\'\i{}nez}}, \bibinfo {author} {\bibfnamefont {J.}~\bibnamefont {Hern\'andez-Garc\'\i{}a}}, \bibinfo {author} {\bibfnamefont {J.}~\bibnamefont {L\'opez-Pav\'on}}, \bibinfo {author} {\bibfnamefont {X.}~\bibnamefont {Marcano}}, \ and\ \bibinfo {author} {\bibfnamefont {D.}~\bibnamefont {Naredo-Tuero}},\ }\href {\doibase 10.1007/JHEP08(2023)030} {\bibfield  {journal} {\bibinfo  {journal} {JHEP}\ }\textbf {\bibinfo {volume} {08}},\ \bibinfo {pages} {030} (\bibinfo {year} {2023})},\ \Eprint {http://arxiv.org/abs/2306.01040} {arXiv:2306.01040 [hep-ph]} \BibitemShut {NoStop}%
\bibitem [{\citenamefont {Gronau}\ \emph {et~al.}(1984)\citenamefont {Gronau}, \citenamefont {Leung},\ and\ \citenamefont {Rosner}}]{Gronau:1984ct}%
  \BibitemOpen
  \bibfield  {author} {\bibinfo {author} {\bibfnamefont {M.}~\bibnamefont {Gronau}}, \bibinfo {author} {\bibfnamefont {C.~N.}\ \bibnamefont {Leung}}, \ and\ \bibinfo {author} {\bibfnamefont {J.~L.}\ \bibnamefont {Rosner}},\ }\href {\doibase 10.1103/PhysRevD.29.2539} {\bibfield  {journal} {\bibinfo  {journal} {Phys. Rev. D}\ }\textbf {\bibinfo {volume} {29}},\ \bibinfo {pages} {2539} (\bibinfo {year} {1984})}\BibitemShut {NoStop}%
\bibitem [{\citenamefont {Nardi}\ \emph {et~al.}(1994)\citenamefont {Nardi}, \citenamefont {Roulet},\ and\ \citenamefont {Tommasini}}]{Nardi:1994iv}%
  \BibitemOpen
  \bibfield  {author} {\bibinfo {author} {\bibfnamefont {E.}~\bibnamefont {Nardi}}, \bibinfo {author} {\bibfnamefont {E.}~\bibnamefont {Roulet}}, \ and\ \bibinfo {author} {\bibfnamefont {D.}~\bibnamefont {Tommasini}},\ }\href {\doibase 10.1016/0370-2693(94)90736-6} {\bibfield  {journal} {\bibinfo  {journal} {Phys. Lett. B}\ }\textbf {\bibinfo {volume} {327}},\ \bibinfo {pages} {319} (\bibinfo {year} {1994})},\ \Eprint {http://arxiv.org/abs/hep-ph/9402224} {arXiv:hep-ph/9402224} \BibitemShut {NoStop}%
\bibitem [{\citenamefont {Atre}\ \emph {et~al.}(2009)\citenamefont {Atre}, \citenamefont {Han}, \citenamefont {Pascoli},\ and\ \citenamefont {Zhang}}]{Atre:2009rg}%
  \BibitemOpen
  \bibfield  {author} {\bibinfo {author} {\bibfnamefont {A.}~\bibnamefont {Atre}}, \bibinfo {author} {\bibfnamefont {T.}~\bibnamefont {Han}}, \bibinfo {author} {\bibfnamefont {S.}~\bibnamefont {Pascoli}}, \ and\ \bibinfo {author} {\bibfnamefont {B.}~\bibnamefont {Zhang}},\ }\href {\doibase 10.1088/1126-6708/2009/05/030} {\bibfield  {journal} {\bibinfo  {journal} {JHEP}\ }\textbf {\bibinfo {volume} {05}},\ \bibinfo {pages} {030} (\bibinfo {year} {2009})},\ \Eprint {http://arxiv.org/abs/0901.3589} {arXiv:0901.3589 [hep-ph]} \BibitemShut {NoStop}%
\bibitem [{\citenamefont {Fernandez-Martinez}\ \emph {et~al.}(2016)\citenamefont {Fernandez-Martinez}, \citenamefont {Hernandez-Garcia},\ and\ \citenamefont {Lopez-Pavon}}]{Fernandez-Martinez:2016lgt}%
  \BibitemOpen
  \bibfield  {author} {\bibinfo {author} {\bibfnamefont {E.}~\bibnamefont {Fernandez-Martinez}}, \bibinfo {author} {\bibfnamefont {J.}~\bibnamefont {Hernandez-Garcia}}, \ and\ \bibinfo {author} {\bibfnamefont {J.}~\bibnamefont {Lopez-Pavon}},\ }\href {\doibase 10.1007/JHEP08(2016)033} {\bibfield  {journal} {\bibinfo  {journal} {JHEP}\ }\textbf {\bibinfo {volume} {08}},\ \bibinfo {pages} {033} (\bibinfo {year} {2016})},\ \Eprint {http://arxiv.org/abs/1605.08774} {arXiv:1605.08774 [hep-ph]} \BibitemShut {NoStop}%
\bibitem [{\citenamefont {He}\ \emph {et~al.}(2003)\citenamefont {He}, \citenamefont {Cheng},\ and\ \citenamefont {Li}}]{He:2002pva}%
  \BibitemOpen
  \bibfield  {author} {\bibinfo {author} {\bibfnamefont {B.}~\bibnamefont {He}}, \bibinfo {author} {\bibfnamefont {T.~P.}\ \bibnamefont {Cheng}}, \ and\ \bibinfo {author} {\bibfnamefont {L.-F.}\ \bibnamefont {Li}},\ }\href {\doibase 10.1016/S0370-2693(02)03258-6} {\bibfield  {journal} {\bibinfo  {journal} {Phys. Lett. B}\ }\textbf {\bibinfo {volume} {553}},\ \bibinfo {pages} {277} (\bibinfo {year} {2003})},\ \Eprint {http://arxiv.org/abs/hep-ph/0209175} {arXiv:hep-ph/0209175} \BibitemShut {NoStop}%
\bibitem [{\citenamefont {Ilakovac}\ and\ \citenamefont {Pilaftsis}(1995)}]{Ilakovac:1994kj}%
  \BibitemOpen
  \bibfield  {author} {\bibinfo {author} {\bibfnamefont {A.}~\bibnamefont {Ilakovac}}\ and\ \bibinfo {author} {\bibfnamefont {A.}~\bibnamefont {Pilaftsis}},\ }\href {\doibase 10.1016/0550-3213(94)00567-X} {\bibfield  {journal} {\bibinfo  {journal} {Nucl. Phys. B}\ }\textbf {\bibinfo {volume} {437}},\ \bibinfo {pages} {491} (\bibinfo {year} {1995})},\ \Eprint {http://arxiv.org/abs/hep-ph/9403398} {arXiv:hep-ph/9403398} \BibitemShut {NoStop}%
\bibitem [{\citenamefont {Hern\'andez-Tom\'e}\ \emph {et~al.}(2020)\citenamefont {Hern\'andez-Tom\'e}, \citenamefont {Illana}, \citenamefont {Masip}, \citenamefont {L\'opez~Castro},\ and\ \citenamefont {Roig}}]{Hernandez-Tome:2019lkb}%
  \BibitemOpen
  \bibfield  {author} {\bibinfo {author} {\bibfnamefont {G.}~\bibnamefont {Hern\'andez-Tom\'e}}, \bibinfo {author} {\bibfnamefont {J.~I.}\ \bibnamefont {Illana}}, \bibinfo {author} {\bibfnamefont {M.}~\bibnamefont {Masip}}, \bibinfo {author} {\bibfnamefont {G.}~\bibnamefont {L\'opez~Castro}}, \ and\ \bibinfo {author} {\bibfnamefont {P.}~\bibnamefont {Roig}},\ }\href {\doibase 10.1103/PhysRevD.101.075020} {\bibfield  {journal} {\bibinfo  {journal} {Phys. Rev. D}\ }\textbf {\bibinfo {volume} {101}},\ \bibinfo {pages} {075020} (\bibinfo {year} {2020})},\ \Eprint {http://arxiv.org/abs/1912.13327} {arXiv:1912.13327 [hep-ph]} \BibitemShut {NoStop}%
\bibitem [{\citenamefont {Adam}\ \emph {et~al.}(2013)\citenamefont {Adam} \emph {et~al.}}]{MEG:2013oxv}%
  \BibitemOpen
  \bibfield  {author} {\bibinfo {author} {\bibfnamefont {J.}~\bibnamefont {Adam}} \emph {et~al.} (\bibinfo {collaboration} {MEG}),\ }\href {\doibase 10.1103/PhysRevLett.110.201801} {\bibfield  {journal} {\bibinfo  {journal} {Phys. Rev. Lett.}\ }\textbf {\bibinfo {volume} {110}},\ \bibinfo {pages} {201801} (\bibinfo {year} {2013})},\ \Eprint {http://arxiv.org/abs/1303.0754} {arXiv:1303.0754 [hep-ex]} \BibitemShut {NoStop}%
\bibitem [{\citenamefont {Baldini}\ \emph {et~al.}(2018)\citenamefont {Baldini} \emph {et~al.}}]{MEGII:2018kmf}%
  \BibitemOpen
  \bibfield  {author} {\bibinfo {author} {\bibfnamefont {A.~M.}\ \bibnamefont {Baldini}} \emph {et~al.} (\bibinfo {collaboration} {MEG II}),\ }\href {\doibase 10.1140/epjc/s10052-018-5845-6} {\bibfield  {journal} {\bibinfo  {journal} {Eur. Phys. J. C}\ }\textbf {\bibinfo {volume} {78}},\ \bibinfo {pages} {380} (\bibinfo {year} {2018})},\ \Eprint {http://arxiv.org/abs/1801.04688} {arXiv:1801.04688 [physics.ins-det]} \BibitemShut {NoStop}%
\bibitem [{\citenamefont {Aubert}\ \emph {et~al.}(2010)\citenamefont {Aubert} \emph {et~al.}}]{BaBar:2009hkt}%
  \BibitemOpen
  \bibfield  {author} {\bibinfo {author} {\bibfnamefont {B.}~\bibnamefont {Aubert}} \emph {et~al.} (\bibinfo {collaboration} {BaBar}),\ }\href {\doibase 10.1103/PhysRevLett.104.021802} {\bibfield  {journal} {\bibinfo  {journal} {Phys. Rev. Lett.}\ }\textbf {\bibinfo {volume} {104}},\ \bibinfo {pages} {021802} (\bibinfo {year} {2010})},\ \Eprint {http://arxiv.org/abs/0908.2381} {arXiv:0908.2381 [hep-ex]} \BibitemShut {NoStop}%
\bibitem [{\citenamefont {Altmannshofer}\ \emph {et~al.}(2019)\citenamefont {Altmannshofer} \emph {et~al.}}]{Belle-II:2018jsg}%
  \BibitemOpen
  \bibfield  {author} {\bibinfo {author} {\bibfnamefont {W.}~\bibnamefont {Altmannshofer}} \emph {et~al.} (\bibinfo {collaboration} {Belle-II}),\ }\href {\doibase 10.1093/ptep/ptz106} {\bibfield  {journal} {\bibinfo  {journal} {PTEP}\ }\textbf {\bibinfo {volume} {2019}},\ \bibinfo {pages} {123C01} (\bibinfo {year} {2019})},\ \bibinfo {note} {[Erratum: PTEP 2020, 029201 (2020)]},\ \Eprint {http://arxiv.org/abs/1808.10567} {arXiv:1808.10567 [hep-ex]} \BibitemShut {NoStop}%
\end{thebibliography}%
\end{document}